\documentclass[11pt,preprint]{aastex}
\newcommand{\kms}{km s$^{-1}$}
\newcommand{\zabs}{$z_{\rm abs}$}

\newcommand{\lya}{Ly$\alpha$\ }
\newcommand{\lyb}{Ly$\beta$\ }

\slugcomment{{\sc The Astrophysical Journal}, 575:697-711, 2002 Aug. 20}
\lefthead{TRIPP et al.}
\righthead{VIRGO ABSORBER METALLICITIES}
\begin{document}

\title{The Heavy Element Enrichment of \lya Clouds in the 
Virgo Supercluster\altaffilmark{1}}

\author{T. M. Tripp,\altaffilmark{2,3} E. B. 
Jenkins,\altaffilmark{2}, G. M. 
Williger,\altaffilmark{4,5,6} S. R. Heap,\altaffilmark{4}, 
C. W. Bowers,\altaffilmark{4} A. C. Danks ,\altaffilmark{4} 
R. Dav\'{e},\altaffilmark{7,8} R. F. Green,\altaffilmark{5} 
T. R. Gull,\altaffilmark{4} C. L. Joseph,\altaffilmark{9} 
M. E. Kaiser,\altaffilmark{6} D. Lindler,\altaffilmark{4} 
R. J. Weymann,\altaffilmark{10} and B. E. 
Woodgate\altaffilmark{4}}

\altaffiltext{1}{ Based on observations with the NASA/ESA 
{\it Hubble Space Telescope}, obtained at the Space 
Telescope Science Institute, which is operated by the 
Association of Universities for Research in Astronomy, 
Inc., under NASA contract NAS 5-26555.}

\altaffiltext{2}{Princeton University Observatory, 
Peyton Hall, Princeton, NJ 08544}
\altaffiltext{3}{Electronic mail: tripp@astro.princeton.edu}

\altaffiltext{4}{NASA Goddard Space Flight Center, Code 681, Greenbelt, 
MD 20771}

\altaffiltext{5}{National Optical Astronomy Observatories, Tucson, AZ 
85726}

\altaffiltext{6}{Department of Physics and Astronomy, Johns Hopkins 
University, 3400 North Charles Street, Baltimore, MD 21218}

\altaffiltext{7}{Steward Observatory, University of Arizona, 933 North 
Cherry Avenue, Tucson, AZ 85721}

\altaffiltext{8}{Hubble Fellow.}

\altaffiltext{9}{Department of Physics and Astronomy, Rutgers 
University, New Brunswick, NJ 08855}

\altaffiltext{10}{Observatories of the Carnegie Institution of 
Washington, 813 Santa Barbara Street, Pasadena, CA 91101-1292}

\begin{abstract}
Using high signal-to-noise echelle spectra of 3C 273 
obtained with the Space Telescope Imaging Spectrograph 
(resolution = 7 \kms\ FWHM), we constrain the metallicities 
of two \lya clouds in the vicinity of the Virgo cluster. We 
detect \ion{C}{2}, \ion{Si}{2}, and \ion{Si}{3} absorption 
lines in the \lya absorber at \zabs\ = 0.00530. Previous 
observations with the {\it Far Ultraviolet Spectroscopic 
Explorer} have revealed Ly$\beta - $Ly$\theta$ absorption 
lines at the same redshift, thereby accurately constraining 
the \ion{H}{1} column density. We model the ionization of 
the gas and derive [C/H] = $-1.2^{+0.3}_{-0.2}$, [Si/C] = 
$+0.2\pm 0.1$, and log $n_{\rm H} = -2.8\pm 0.3$. The model 
implies a small absorber thickness, $\sim 70$ pc, and 
thermal pressure $p/k \approx 40$ cm$^{-3}$ K. It is most 
likely that the absorber is pressure confined by an 
external medium because gravitational confinement would 
require a very high ratio of dark matter to baryonic 
matter. Based on a sample of Milky Way sight lines in which 
carbon and silicon abundances have been reliably measured 
in the same interstellar cloud (including new measurements 
presented herein), we argue that it is unlikely that the 
overabundance of Si relative to C is due to depletion onto 
dust grains. Instead, this probably indicates that the gas 
has been predominately enriched by ejecta from Type II 
supernovae. Such enrichment is most plausibly provided by 
an unbound galactic wind, given the absence of known 
galaxies within a projected distance of 100 kpc and the 
presence of galaxies capable of driving a wind at larger 
distances (e.g., \ion{H}{1} 1225+01). Such processes have 
been invoked to explain the observed abundances in the hot, 
X-ray emitting gas in Virgo. However, the sight line to 3C 
273 is more than 10$^{\circ}$ away from the X-ray emission 
region. We also constrain the metallicity and physical 
conditions of the Virgo absorber at \zabs\ = 0.00337 in the 
spectrum of 3C 273 based on detections of \ion{O}{6} and 
\ion{H}{1} and an upper limit on \ion{C}{4}. If this 
absorber is collisionally ionized, the 
\ion{O}{6}/\ion{C}{4} limit requires $T \gtrsim 10^{5.3}$ K 
in the \ion{O}{6}-bearing gas. For either collisional 
ionization or photoionization, we find that [O/H] $\gtrsim 
-2.0$ at \zabs\ = 0.00337.
\end{abstract}


\section{Introduction}

The heavy element enrichment of the intergalactic medium 
(IGM) provides a valuable constraint on physical processes 
that play critical roles in the evolution of galaxies. 
Elements heavier than boron are synthesized in stars 
(presumably mostly located within galaxies), and 
understanding how these elements are removed from galaxies 
and transported into the IGM will likely yield insights on 
galaxy evolution. Is the IGM enriched by supernova-driven 
galactic winds, dynamical processes such as tidal or 
ram-pressure stripping, or all of these mechanisms? How 
does this affect/regulate star formation in galaxies? IGM 
enrichment also has cosmological implications. For example, 
the processes which deposit heavy elements in the IGM will 
also deposit energy, and this ``feedback'' can have 
important effects on cosmological structure growth (e.g., 
Kaiser 1991; Ponman, Cannon, \& Navarro 1999; Cen \& Bryan 
2001; Voit \& Bryan 2001). 

IGM enrichment has been modeled using methods ranging from 
semi-analytic calculations to hydrodynamic simulations, and 
a variety of theoretical predictions have been made 
regarding the heavy element content of intergalactic gas at 
both high and low redshifts (e.g., Cen \& Ostriker 1999b; 
Ferrara, Pettini, \& Shchekinov 2000; Aguirre et al. 2001). 
These predictions can be tested using QSO absorption lines 
to estimate the metallicity of the IGM in a variety of 
contexts. Observations of QSO absorption-line systems have 
shown that the high-redshift IGM is remarkably-widely 
enriched with metals (Cowie et al. 1995; Tytler et al. 
1995), and statistical analyses have indicated that this 
enrichment extends even to the absorbers with the lowest 
\ion{H}{1} column densities (Ellison et al. 2000; Schaye et 
al. 2000). 

Recently, high signal-to-noise (S/N) UV spectroscopy of 3C 
273 with the Space Telescope Imaging Spectrograph (STIS) 
and the {\it Far Ultraviolet Spectroscopic Explorer (FUSE)} 
has provided an opportunity to probe the IGM enrichment 
within a particularly interesting environment: the Virgo 
cluster/supercluster. In this paper we derive constraints 
on the heavy element abundances in two \lya clouds in the 
immediate vicinity of this cluster. The sight line to 3C 
273 extends through the Virgo cluster ``Southern 
Extension'' (e.g., Binggeli, Popescu, \& Tammann 1993), 
although the projected distance from the sight line to M87 
and the X-ray region is substantial ($> 3$ Mpc). In the 3C 
273 spectrum, there are three \lya clouds at Virgo 
redshifts; in this paper we are mainly interested in the 
Virgo absorbers at \zabs\ = 0.00337 and 0.00530.  The 
connections between these Virgo absorption systems and 
nearby galaxies have been extensively studied (see \S 4.3), 
but measurements of the metallicities of these \lya 
absorbers have not been previously obtained. It is now 
possible to estimate the metal abundances in these Virgo 
\lya clouds due to two observational advances. First, 
Sembach et al. (2001) have presented a surprising result 
from the {\it FUSE} observation: the Virgo absorber at 
\zabs\ = 0.00530 is detected in {\it seven} higher Lyman 
series lines (\lyb - Ly$\theta$). This clearly establishes 
that the \ion{H}{1} column is substantially higher than 
previously thought; from a curve-of-growth (COG) analysis, 
Sembach et al. (2001) derive log N(\ion{H}{1}) = 
15.85$^{+0.10}_{-0.08}$, an increase by a factor of more 
than 40 over the previous estimate based on fitting the 
profile of \lya only.\footnote{Using {\it ORFEUS II} 
observations of the Ly$\beta$ line at this redshift, 
Hurwitz et al. (1998) provided the first evidence that the 
column density of this system was substantially higher than 
expected based on profile fitting of the Ly$\alpha$ 
transition alone. However, with only Ly$\alpha$ and 
Ly$\beta$, they were unable to tightly constrain $N$({\sc H 
i}) and $b$ (see their Figure 4).} More importantly, the 
\ion{H}{1} column density and Doppler parameter are 
accurately constrained from the COG. Second, the high 
signal-to-noise STIS observations at 7 \kms\ resolution 
have enabled the detection of very weak lines in the 3C 273 
spectrum, including weak lines of \ion{C}{2}, \ion{Si}{2}, 
and \ion{Si}{3} at \zabs\ = 0.00530. In this paper, we use 
these weak metal lines and the new \ion{H}{1} measurement 
to explore the heavy element enrichment of this Virgo \lya 
system at \zabs\ = 0.00530. We also constrain the 
metallicity of the 
lower-redshift Virgo \lya cloud (\zabs\ = 0.00337) based on 
the detection of \ion{O}{6} and the absence of \ion{C}{4}. 

The paper is organized as follows. In \S 2 we present the 
new STIS observations and absorption line measurements, and 
we briefly comment on the well-known galaxy structures at 
Virgo redshifts in the vicinity of the 3C 273 sight line. 
In \S 3 we model the ionization of the gas and derive 
constraints on the metallicity, density, and temperature of 
the \lya absorbers. We discuss the implications of the 
measurements and the nature of the absorbing gas in \S 4, 
and we summarize the main results in \S 5. In the Appendix 
we provide additional details on new Galactic ISM 
measurements to supplement the analysis in \S 4. Throughout 
the paper, we report heliocentric wavelengths and 
redshifts,\footnote{In the direction of 3C 273, $v_{\rm 
LSR} = v_{\odot} + 2.3$ \kms\ assuming the standard 
definition of the Local Standard of Rest (Kerr \& Lynden-
Bell 1986).} and we assume $H_{0} = 75$ \kms\ Mpc$^{-1}$ 
and $q_{0} = 0$.
\clearpage

\section{Observations and Measurements}

Observations of 3C 273 were obtained with the E140M echelle 
mode of STIS, using the $0\farcs 2 \times 0\farcs 2$ 
aperture, on 2000 May 2 and 2000 June 21-22. This mode 
provides a resolution of 7 \kms\ (FWHM) and records the 
spectrum between 1150 and 1700 \AA\ with only a few small 
gaps between orders at $\lambda >$ 1615 \AA\ (Kimble et al. 
1998; Woodgate et al. 1998). The total integration time was 
18.67 ksec, and this resulted in signal-to-noise ratios 
ranging from $\sim$6 to 66 per resolution element. The data 
reduction followed standard procedures, including the 
scattered light correction developed by the STIS 
Investigation Definition Team. 

\begin{figure}
\plotone{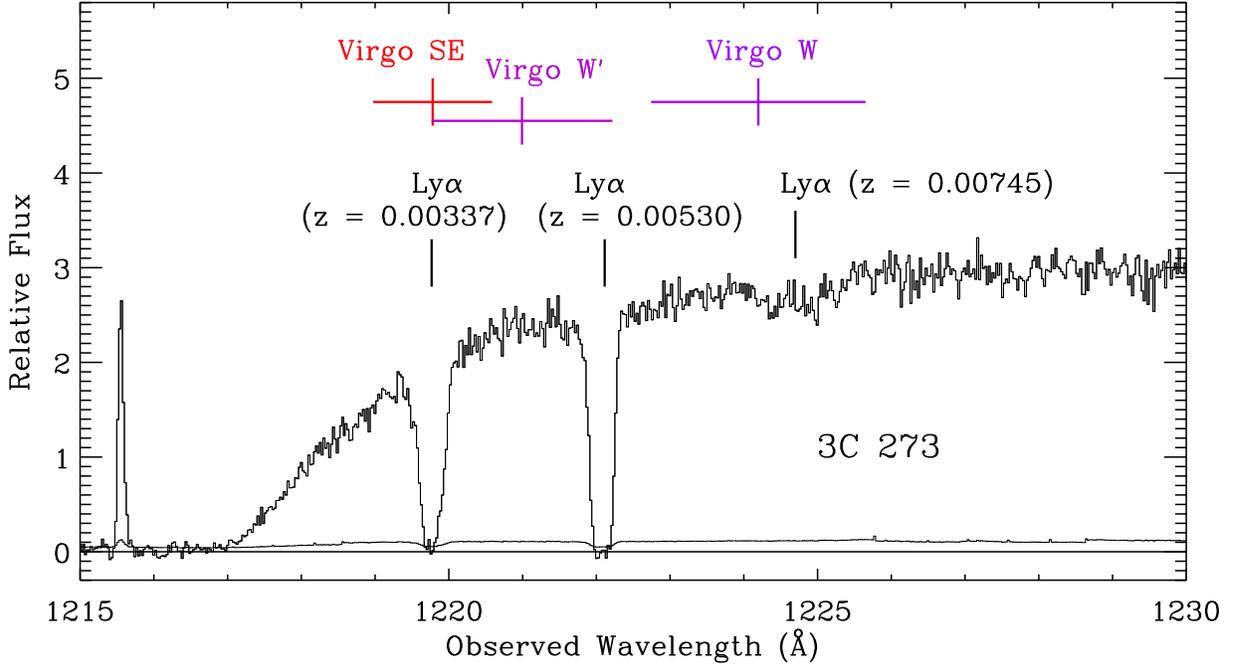}
\caption[]{Portion of the STIS echelle spectrum of 3C 273 
plotted versus observed heliocentric wavelength. In this 
figure, the spectrum has been binned (two pixels into one) 
for display purposes only (all other figures and 
measurements make use of the full-resolution, unbinned 
data). The three \lya clouds within the Virgo cluster 
region at \zabs\ = 0.00337, 0.00530, and 0.00745 are 
indicated. The redshifted \lya wavelengths corresponding to 
the mean velocities of the ``Southern Extension'', W, and 
W' galaxy clouds in Virgo (from Binggeli et al. 1993) are 
also indicated; the horizontal bars indicate the velocity 
dispersions of these Virgo structures. The solid line near 
zero is the 1$\sigma$ flux uncertainty, adjusted to reflect 
the binning. The broad absorption feature centered at 1216 
\AA\ is the damped \lya line due to \ion{H}{1} in the Milky 
Way ISM, and the spike is the geocoronal \lya emission 
line.\label{virgolya}}
\end{figure}

A selected portion of the final spectrum is shown in 
Figure~\ref{virgolya}. This portion of the spectrum spans 
the wavelength range of \lya lines at redshifts within the 
Virgo cluster/supercluster. Previous {\it HST} studies have 
detected two \lya clouds within the Virgo cluster at high 
significance levels at \zabs\ = 0.003 and 0.005, but a 
third \lya absorber at \zabs\ = 0.007 was reported in some 
studies but not confirmed (see \S 3.3.1 of Morris et al. 
1993). For example, Brandt et al. (1997) do not list a line 
at \zabs\ = 0.007 in their intermediate-resolution GHRS 
atlas of 3C 273. However, Penton et al. (2000) claim a 
significant detection of this line using GHRS/G160M data, 
and a weak feature is evident at the right wavelength in 
the Brandt et al. spectrum. The new STIS spectrum settles 
this question: a broad feature with rest-frame equivalent 
width\footnote{Rest-frame equivalent width $W_{\rm r} = 
W_{\rm obs}/(1 + z_{\rm abs})$, where $W_{\rm obs}$ is the 
observed equivalent width.} $W_{\rm r} = 94 \pm 14$ m\AA\ 
is detected at \zabs\ = 0.00745.

\begin{figure}
\plotone{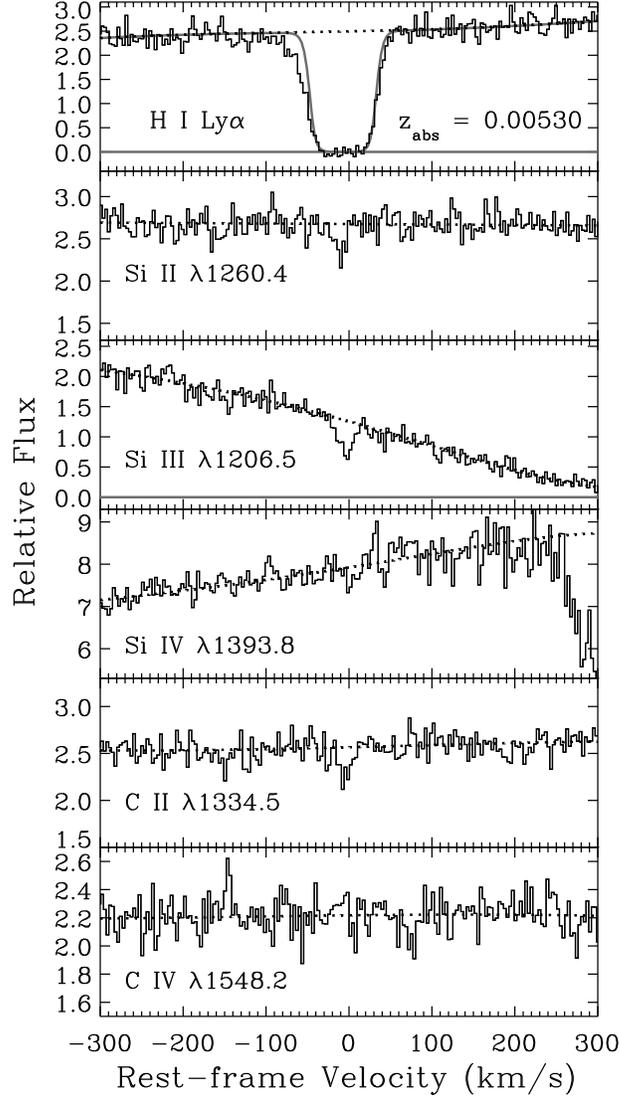}
\caption[]{\footnotesize Absorption profiles of the H~I 
Ly$\alpha$, Si~II $\lambda 1260.4$, 
Si~III $\lambda 1206.5$, and C~II $\lambda 
1334.5$ lines associated with the \lya absorber at \zabs\ = 
0.00530 in the 3C 273 spectrum, plotted vs. rest-frame 
velocity where $v = 0$ \kms\ at \zabs\ = 0.00530. The 
spectral regions of the Si~IV $\lambda 1393.8$ and 
C~IV $\lambda 1548.2$ lines, which are not detected, 
are also shown. The dotted lines show the continua adopted 
for the measurements and upper limits listed in 
Table~\ref{lineprop}. Note that the expected wavelength of 
the Si~IV line is near the region affected by the FUV 
MAMA repeller wire, and the feature at $v \sim$ 300 \kms\ 
is an artifact due to the repeller which was not adequately 
removed by the flat field. The smooth solid curve in the 
top panel shows a theoretical \lya profile with log 
$N$(H~I) = 15.85 and $b$ = 16.1 \kms , as derived by 
Sembach et al. (2001) from a curve-of-growth analysis of 
{\it FUSE} observations of 3C 273. The line spread function 
from the STIS Handbook has been assumed to produce this 
theoretical profile, and its velocity centroid has been 
adjusted to best fit the observed line. The observed \lya 
profile is $\sim$ 50 m\AA\ stronger than expected; the 
excess absorption is evident primarily in the blue wing of 
the line.\label{stack}}
\end{figure}

\begin{figure}
\plotone{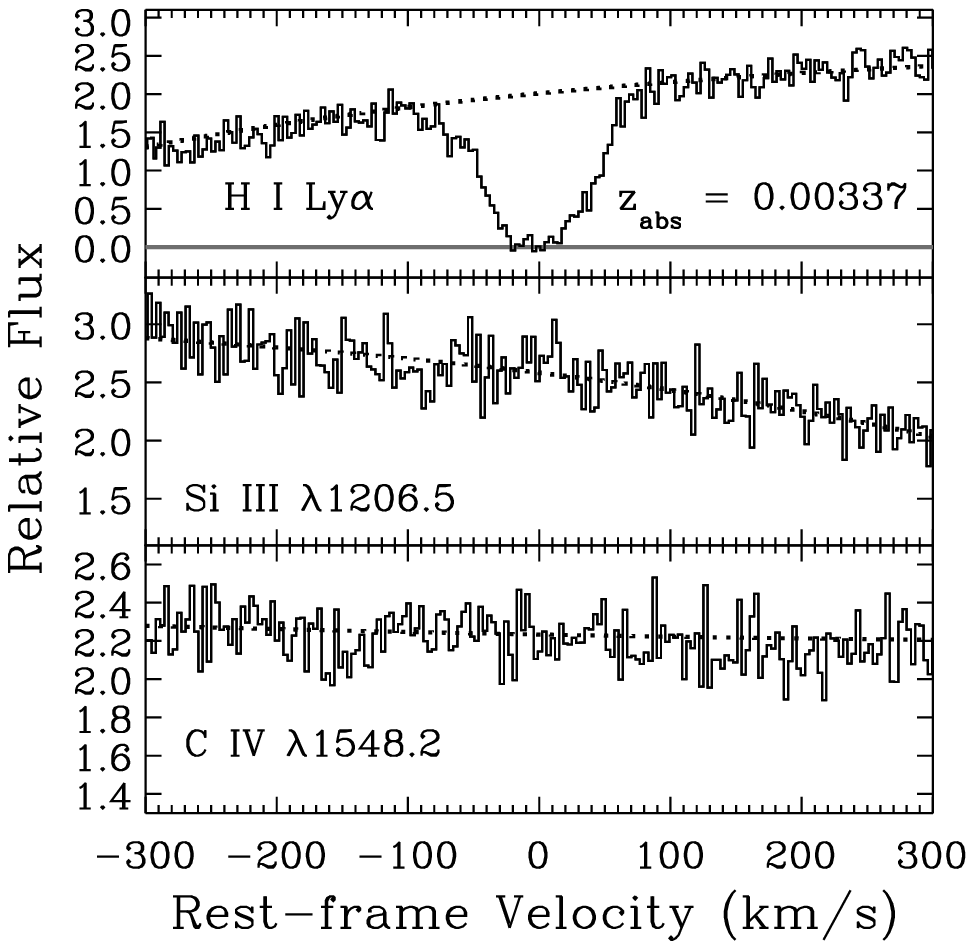}
\caption[]{Absorption profile of the \ion{H}{1} Ly$\alpha$ 
line at \zabs\ = 0.00337 in the direction of 3C 273, 
plotted versus rest-frame velocity where $v = 0$ at \zabs\ 
= 0.00337. The spectral regions of the undetected 
\ion{Si}{3} $\lambda$1206.5 and \ion{C}{4} $\lambda$1548.2 
lines are also shown.\label{stack003}}
\end{figure}

Figure~\ref{virgolya} also indicates the expected 
wavelengths of \lya lines at the mean redshifts of several 
Virgo galaxy structures near the 3C 273 sight line. It has 
long been recognized that there are several large 
``clouds'' of galaxies in the immediate vicinity of the 
Virgo cluster that have distinct kinematics and 
morphologies (e.g., de Vaucouleurs 1961). According to 
Binggeli, Popescu, \& Tammann (1993), the 3C 273 sight line 
pierces the structures known as the Virgo ``Southern 
Extension'' ($<v> = 1012 \pm 51$ \kms , velocity dispersion 
$\sigma = 198$ \kms ) and the Virgo W cloud ($<v> = 2099 
\pm 41$ \kms , $\sigma$ = 356 \kms ). The Virgo W' cloud is 
also in this region ($<v> = 1310 \pm 80$ \kms , $\sigma$ = 
301 \kms ). The relationships between these structures and 
the Virgo cluster proper are not entirely clear. As noted, 
our pencil beam to 3C 273 is at a substantial projected 
distance from the Virgo X-ray emission. At any rate, the 
sight line probes a structure that contains a large number 
of galaxies in the general vicinity of the Virgo cluster. 
For convenience, we simply refer to this region as the 
Virgo supercluster in this paper.

Figure~\ref{stack} shows the absorption profiles of lines 
detected in the \zabs\ = 0.00530 absorber including 
\ion{H}{1} Ly$\alpha$, \ion{Si}{2} $\lambda$1260.4, 
\ion{Si}{3} $\lambda$1206.5, and \ion{C}{2} 
$\lambda$1334.5, as well as the regions of the spectrum 
covering the undetected \ion{Si}{4} $\lambda$1393.8 and 
\ion{C}{4} $\lambda$1548.2 lines. We note the presence of a 
very weak feature near the expected wavelength of the 
\ion{Si}{4} transition. However, the significance of this 
feature is much less than $3\sigma$, and furthermore this 
spectral region is partially obscured by the field 
electrode of the FUV-MAMA detector, the ``repeller wire'' 
(Woodgate et al. 1998). In the majority of the 3C 273 
observations, the repeller introduces significant artifacts 
into the spectrum which are not adequately removed by the flat 
field. Sembach et al. (2001) do not report significant 
detection of any metals at \zabs\ = 0.00530 in the {\it 
FUSE} 3C 273 spectrum, but the stronger transition of the 
\ion{O}{6} doublet is blended with strong Milky Way 
absorption lines. Similarly, the strong \ion{C}{3} 
$\lambda$977.0 line falls close to a Galactic H$_{2}$ 
absorption line at this redshift.

The only line detected at \zabs\ = 0.00337 in the STIS 
E140M spectrum of 3C 273 is the \lya transition. However, 
Sembach et al. (2001) have also detected Ly$\beta$ and 
\ion{O}{6} $\lambda$1031.9 at this redshift in the {\it 
FUSE} spectrum. Figure~\ref{stack003} shows the \lya\ line 
at \zabs\ = 0.00337 as well as regions of strong metal 
lines that are not detected.

We have measured the absorption-line equivalent widths and 
integrated apparent column densities\footnote{The 
integrated apparent column density is given by $N_{\rm a} = 
(m_{\rm e}c/\pi e^{2})(f\lambda )^{-1} \int \tau _{\rm 
a}(v) dv = 3.768 \times 10^{14} (f\lambda )^{-1} \int {\rm 
ln}[I_{\rm c}(v)/I(v)] dv$, where $f$ is the oscillator 
strength, $\lambda$ is the wavelength (the numerical 
coefficient in the latter expression assumes $\lambda$ is 
in \AA ), $I(v)$ is the observed line intensity and $I_{\rm 
c} (v)$ is the estimated continuum intensity at velocity 
$v$; see Savage \& Sembach (1991) for further details.} 
using the techniques of Sembach \& Savage (1992), which 
include in the overall error evaluation the uncertainty in 
the level and curvature of the continuum and a 2\% flux 
zero point uncertainty, as well as the usual 
photon-counting uncertainty. The equivalent widths and 
column densities of the \zabs\ = 0.00530 absorption system 
are summarized in Table~\ref{lineprop} along with upper 
limits on undetected species of interest. Equivalent 
widths, column densities, and upper limits from the \zabs\ 
= 0.00337 absorber are listed in Table~\ref{z003prop}.

Based on previous GHRS observations of 3C 273, Sembach et 
al. (2000) report that the equivalent width of the \lya 
line at \zabs\ = 0.00530 is $\sim$40 m\AA\ stronger than 
predicted by their COG analysis of the \lyb - Ly$\theta$ 
lines at this redshift. As shown in the top panel of 
Figure~\ref{stack}, we confirm this result with the higher 
resolution STIS data. The excess equivalent width in the 
STIS spectrum is $\sim$50 m\AA\ and is most evident on the 
blue side of the profile. While this establishes that 
multiple components contribute to the \ion{H}{1} 
absorption, the additional component in the \lya profile is 
probably too weak to significantly affect the lines higher 
in the series, and Sembach et al. show that including or 
excluding \lya does not change $N$(\ion{H}{1}) and $b$ 
significantly (see their Table 4). We discuss the 
implications of the \lya profiles below.

\begin{deluxetable}{lcccccc}
\footnotesize
\tablewidth{0pc}
\tablecaption{Equivalent Widths and Integrated Column 
Densities of the Virgo Ly$\alpha$ Absorber at $z_{\rm abs}$ 
= 0.00530\tablenotemark{a}\label{lineprop}}
\tablehead{Species & $\lambda _{0}$\tablenotemark{b} & 
$f$\tablenotemark{b} & $W_{\rm r}$\tablenotemark{c} & 
$<v>$\tablenotemark{d} & log $N$ & S/N\tablenotemark{e} \\
 \ & (\AA ) & \ & (m\AA ) & (km s$^{-1}$) & \ & \ }
\startdata
\ion{H}{1}....... & 1215.67 & 0.416 & 377$\pm$12 & $-10\pm 
3$ & $>14.3$ & 23 \\
\ion{Si}{2}...... & 1260.42 & 1.18 & 8.9$\pm$2.6 & $-12\pm 
5$ & 11.76$^{+0.11}_{-0.15}$ & 30 \\
\ion{Si}{3}.... & 1206.50 & 1.67 & 36.8$\pm$6.1 & $-5\pm 2$ 
& 12.33$\pm$0.08 & $11-16$ \\
\ion{Si}{4}....\tablenotemark{f} & 1393.76 & 0.514 & 
$<8.7$\tablenotemark{g} & \nodata & 
$<12.0$\tablenotemark{h} & 27 \\
\ion{C}{2}...... & 1334.53 & 0.127 & 10.3$\pm$2.6 & $-9\pm 
3$ & 12.74$^{+0.10}_{-0.12}$ & 34 \\
\ion{C}{4}..... & 1548.20 & 0.191 & 
$<12.0$\tablenotemark{g} & \nodata & 
$<12.5$\tablenotemark{h} & 25 \\
\ion{Al}{2}...... & 1670.79 & 1.83 & 
$<17.3$\tablenotemark{g} & \nodata & 
$<11.6$\tablenotemark{h} & 17 \\
\ion{Fe}{2}...... & 1608.45 & 0.0580 & $<25.3$ & \nodata & 
$<13.3$ & 16 \\
\ion{O}{1}........ & 1302.17 & 0.0519 & 
$<6.7$\tablenotemark{g} & \nodata & 
$<12.9$\tablenotemark{h} & 34
\enddata
\footnotesize
\tablenotetext{a}{With the exception of Ly$\alpha$, all 
quantitites are integrated from $-30$ to 20 \kms\ where $v 
= 0$ \kms\ at \zabs\ = 0.00530. The Ly$\alpha$ quantities 
are integrated from $-160$ to 70 \kms .}
\tablenotetext{b}{Rest-frame vacuum wavelength and 
oscillator strength from Morton (2002) or Morton (1991).}
\tablenotetext{c}{Rest-frame equivalent width.}
\tablenotetext{d}{Profile-weighted mean velocity of the 
line, $<v> = \int v[1 - I(v)/I_{\rm c}(v)] dv/\int [1 - 
I(v)/I_{\rm c}(v)] dv.$}
\tablenotetext{e}{Signal-to-noise ratio per two-pixel 
resolution element in the continuum near the absorption 
line. The Si~III $\lambda$1206.5 line is located in 
the blue wing of the Milky Way damped \lya profile. Here 
the steep slope of the continuum leads to rapidly 
decreasing S/N with increasing wavelength, so we list the 
S/N range within $\pm$50 km s$^{-1}$ of the line.}
\tablenotetext{f}{The Si~IV $\lambda$1393.8 line is 
located in a region of the spectrum affected by the FUV 
MAMA repeller wire. This introduces features into the 
spectrum which are not adequately removed by the flatfield. 
Furthermore, the line is close to the order edge where 
spurious features are occasionally evident.}
\tablenotetext{g}{$3\sigma$ upper limit.}
\tablenotetext{h}{$3\sigma$ upper limit derived from the 
equivalent width upper limit assuming the linear 
curve-of-growth is appropriate.}
\end{deluxetable}

\begin{deluxetable}{lcccccc}
\footnotesize
\tablewidth{0pc}
\tablecaption{Equivalent Widths and Integrated Column 
Densities of the Virgo Ly$\alpha$ Absorber at $z_{\rm abs}$ 
= 0.00337\tablenotemark{a}\label{z003prop}}
\tablehead{Species & $\lambda _{0}$ & $f$ & $W_{\rm r}$ & 
$<v>$ & log $N$ & S/N \\
 \ & (\AA ) & \ & (m\AA ) & (km s$^{-1}$) & \ & \ }
\startdata
\ion{H}{1}......... & 1215.67 & 0.416 & 389$\pm$12 & $-3\pm 
1$ & $>14.2$ & $17-20$ \\
\ion{Si}{3}.....    & 1206.50 & 1.67  & 
$<$34.5\tablenotemark{b} & \nodata & 
$<$12.2\tablenotemark{b} & 20 \\
\ion{C}{4}...... & 1548.20 & 0.191 & $< 
27.6$\tablenotemark{b} & \nodata & $< 
12.8$\tablenotemark{b} & 24 \\
\ion{O}{6}...... & 1031.93 & 0.0659 & 
25.4$\pm$7.1\tablenotemark{c} & \nodata & $13.32^{+0.13}_{-
0.21}$\tablenotemark{c} & \nodata
\enddata
\tablenotetext{a}{All quantities are integrated from $-100$ 
to 100 \kms\ where $v =$ 0 \kms\ at \zabs\ = 0.00337, 
except the \ion{O}{6} measurements, which are from Sembach 
et al. (2001). See Table~\ref{lineprop} footnotes for 
further details.}
\tablenotetext{b}{3$\sigma$ upper limit.}
\tablenotetext{c}{Equivalent width and column density 
reported by Sembach et al. (2001). The equivalent width is 
the weighted mean of the measurements from the two {\it 
FUSE} channels, and the column density is from their Table 
4.}
\end{deluxetable}
\clearpage

\section{Analysis}

\subsection{\zabs\ = 0.00530}

From the COG analysis of the \lyb - Ly$\theta$ lines at \zabs\ = 
0.00530, Sembach et al. (2001) obtain $b = 16.1\pm 1.1$ 
\kms . Therefore the gas temperature $T \lesssim$ 16,000 K. 
Even though $N$(\ion{H}{1}) has been dramatically revised 
to a larger value in this absorber, with log 
$N$(\ion{H}{1}) = 15.85 the gas is still optically thin 
below the Lyman limit. With this column density, the gas is 
likely to be substantially ionized by the UV background 
radiation from QSOs. Consequently, to derive abundances, 
ionization corrections must be applied. For this purpose, 
we have made use of ionization models constructed with the 
photoionization code CLOUDY (v94.00, Ferland et al. 1998). 
Collisional ionization may also be important, but 
collisional processes are included in CLOUDY, and as we 
shall see, the mean gas temperature in the best CLOUDY 
model is fully consistent with the upper limit implied by 
the $b-$value. Therefore even if collisional ionization is 
dominant, these models should provide reasonable estimates 
of the metal abundances. Throughout this section we shall 
restrict our attention to {\it gas-phase} abundances; we 
discuss how depletion onto dust grains could affect the 
abundances in \S 4.

In the photoionization models, the gas is assumed to be 
adequately approximated by a plane-parallel, constant 
density slab. We adopt the background radiation field at 
$z$ = 0 due to QSOs, as computed by Haardt \& Madau (1996) 
with the mean intensity at 1 Rydberg set to $J_{\nu} = 1 
\times 10^{-23}$ ergs s$^{-1}$ cm$^{-2}$ Hz$^{-1}$ sr$^{-
1}$ (see Shull et al. 1999; Dav\'{e} \& Tripp 2001; Weymann 
et al. 2001 and references therein).\footnote{Scott et al. 
(2002) have recently derived $J_{\nu} = 6.5^{+38}_{-1.6} 
\times 10^{-23}$ ergs s$^{-1}$ cm$^{-2}$ Hz$^{-1}$ sr$^{-
1}$ from the proximity effect in a large number of QSO 
spectra obtained with the Faint Object Spectrograph at 0.03 
$< z <$ 1.0. Given the redshift range to which this 
applies, this is consistent with the expected evolution of 
$J_{\nu}$ and the value we have assumed for $z \approx$ 0.} 
The model column densities primarily depend on the 
ionization parameter $U$ or, equivalently, the gas density 
since $U = n_{\gamma}/n_{\rm H}$ = H ionizing photon 
density/total H number density, and $n_{\gamma}$ is set by 
the assumed shape and intensity of the ionizing radiation 
field. At low metallicities, the column densities scale 
directly with the overall metallicity $Z$. We constrain the 
ionization parameter using the observed 
\ion{Si}{3}/\ion{Si}{2} ratio, but we allow the relative 
abundance of silicon to carbon to vary since the gas may 
not have a solar abundance pattern.

\begin{figure}
\plotone{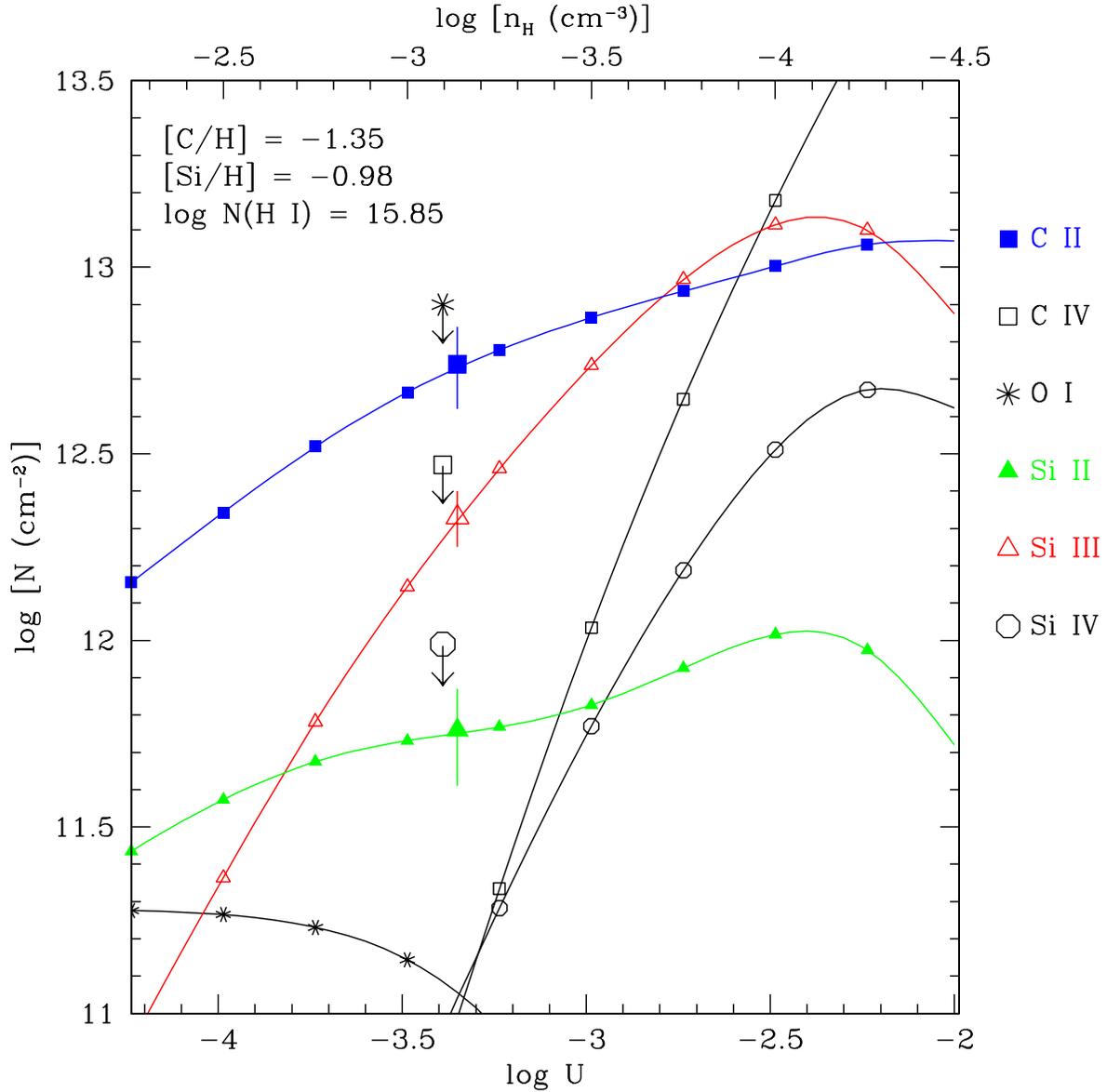}
\caption[]{Heavy element column densities predicted for a 
\lya absorber, photoionized by the UV background from QSOs 
(from Haardt \& Madau 1996) with log $N$(\ion{H}{1}) = 
15.85 and heavy element abundances indicated at upper left, 
compared to the observed column densities at \zabs\ = 
0.00530. The model column densities are indicated with 
solid lines with small symbols while the observed columns 
are shown with larger symbols with 1$\sigma$ error bars (or 
arrows in the case of 3$\sigma$ upper limits). The species 
corresponding to the symbols are shown in the key at the 
right.\label{model}}
\end{figure}

We begin with the most simple model, assuming that the 
\ion{C}{2}, \ion{Si}{2}, and \ion{Si}{3} absorption lines 
arise in the same gas. As we discuss below, there is some 
evidence that these lines do not all occur in the same 
phase, but the evidence is not yet strong enough to rule 
out this possibility. With this assumption, 
Figure~\ref{model} shows the CLOUDY model that provides the 
best simultaneous fit to the observed \ion{C}{2}, 
\ion{Si}{2}, and \ion{Si}{3} column densities. In the usual 
logarithmic notation,\footnote{[X/Y] = log (X/Y) $-$ log 
(X/Y)$_{\odot}$. Solar reference abundances are taken from 
Holweger (2001).} we obtain an excellent fit with the 
following abundances: [C/H] = $-1.4$ and [Si/H] = $-1.0$ 
with $n_{\rm H} \approx 10^{-3.1}$ cm$^{-3}$ (or log $U 
\approx -3.35$). At this density, the model predicts a gas 
temperature of $\sim$ 14,000 K, which is consistent with 
the upper limit on $T$ from the $b-$value obtained by 
Sembach et al. (2001). Figure~\ref{model} also shows that 
the absence of detectable \ion{Si}{4} and \ion{C}{4} 
absorption in this system is fully consistent with the 
model. At log $U \approx -3.35$, the predicted \ion{Si}{4} 
and \ion{C}{4} columns are more than an order of magnitude 
below the observational limits. The model is consistent 
with the other upper limits as well, including the upper 
limit on $N$(\ion{C}{3}) set by Sembach et al. (2001). We 
note that the model is required to produce the observed 
\ion{H}{1} column density, log $N$(\ion{H}{1}) = 15.85, and 
consequently as $U$ is increased and the gas becomes more 
highly ionized, the thickness of the slab is increased to 
match the observed \ion{H}{1} column. This is why 
$N$(\ion{Si}{2}) and $N$(\ion{Si}{3}) both increase with 
increasing $U$ even though the ion fraction of \ion{Si}{2} 
decreases with increasing $U$ over the range shown in 
Figure~\ref{model}.

It is interesting to note that silicon, an $\alpha-$group 
element, is overabundant compared to C by $\sim 0.4$ dex in 
this model. We initially attempted to reproduce the 
observed column densities assuming that the {\it relative} 
heavy element abundances in the Virgo absorber follow the 
solar pattern as well (i. e., [Si/C] = 0.0), but we found 
that this could not be done. This is apparent from a brief 
inspection of Figure~\ref{model}: when the 
\ion{Si}{3}/\ion{Si}{2} ratio is reproduced, the model 
produces too much \ion{C}{2} if [Si/C] = 0.0 (multiphase 
models presented below require a somewhat lower Si 
overabundance, but the best fit in these models still 
yields [Si/C] $\sim$ 0.2). We discuss the implications of 
this result in \S 4.

As shown by Weymann et al. (1995) with regard to H 
ionization, local sources of photoionization, such as 
X-rays from the hot gas in the Virgo cluster or UV flux 
escaping from nearby galaxies, are unlikely to be important 
compared to photoionization by the background UV flux from 
QSOs. We confirm that the heavy element results presented 
above are insensitive to these issues. For example, if we 
include X-ray flux from the hot gas (following \S 3.1.4 in 
Weymann et al. 1995) in the CLOUDY model, we find only 
miniscule changes in the column densities of interest 
compared to Figure~\ref{model}. Similarly, if we assume 
that the radiation field is {\it dominated} by the UV flux 
from stars, we obtain very similar constraints on the heavy 
element abundances. Even lower \ion{Si}{4} and \ion{C}{4} 
column densities are predicted in this case, so the 
observational constraints in Table~\ref{lineprop} are 
easily satisfied. However, if stellar flux were competitive 
with the QSO background, the gas density would be higher; 
the combined flux from the QSO/AGN background and the 
stellar sources leads to a higher photon density, so the 
gas density must also be higher to produce the same 
ionization parameter.

\begin{figure}
\plotone{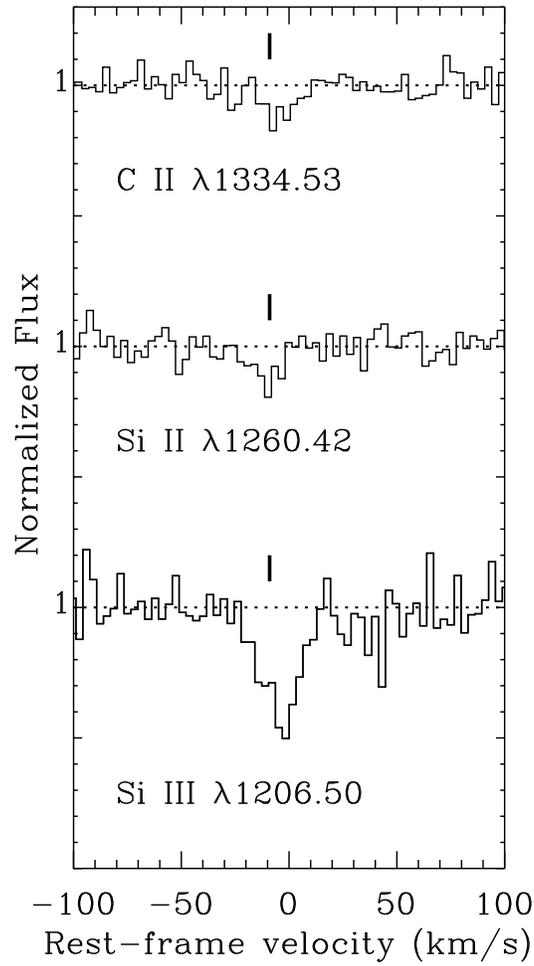}
\caption[]{Continuum-normalized absorption profiles of the 
\ion{C}{2}, \ion{Si}{2}, and \ion{Si}{3} lines at \zabs\ = 
0.00530, plotted on an expanded velocity scale to show the 
apparent differences in the line centroids and component 
structure. For reference, the tick mark indicates a 
velocity of $-9$ \kms .\label{multiphase}}
\end{figure}

We next consider the possibility that the absorption occurs 
in a multiphase medium in which the \ion{C}{2}, 
\ion{Si}{2}, and \ion{Si}{3} lines do not entirely come 
from the same gas. As shown in Figure~\ref{multiphase}, the 
centroids of these lines (see also column 5 in 
Table~\ref{lineprop}) marginally support this hypothesis: 
the \ion{C}{2} and \ion{Si}{2} lines are blueshifted by 
$\sim$5 \kms\ compared to the centroid of the \ion{Si}{3} 
absorption. Also, the \ion{Si}{3} profile appears to show 
two components with the weaker component more closely 
aligned with the \ion{Si}{2} and \ion{C}{2} lines. However, 
the \ion{Si}{2} and \ion{C}{2} lines are very weak and 
therefore could be misleading due to noise. Consequently, 
we conclude that while there are reasonable indications 
that this is a multiphase absorption system, this requires 
confirmation with higher S/N data.

\begin{figure}
\plotone{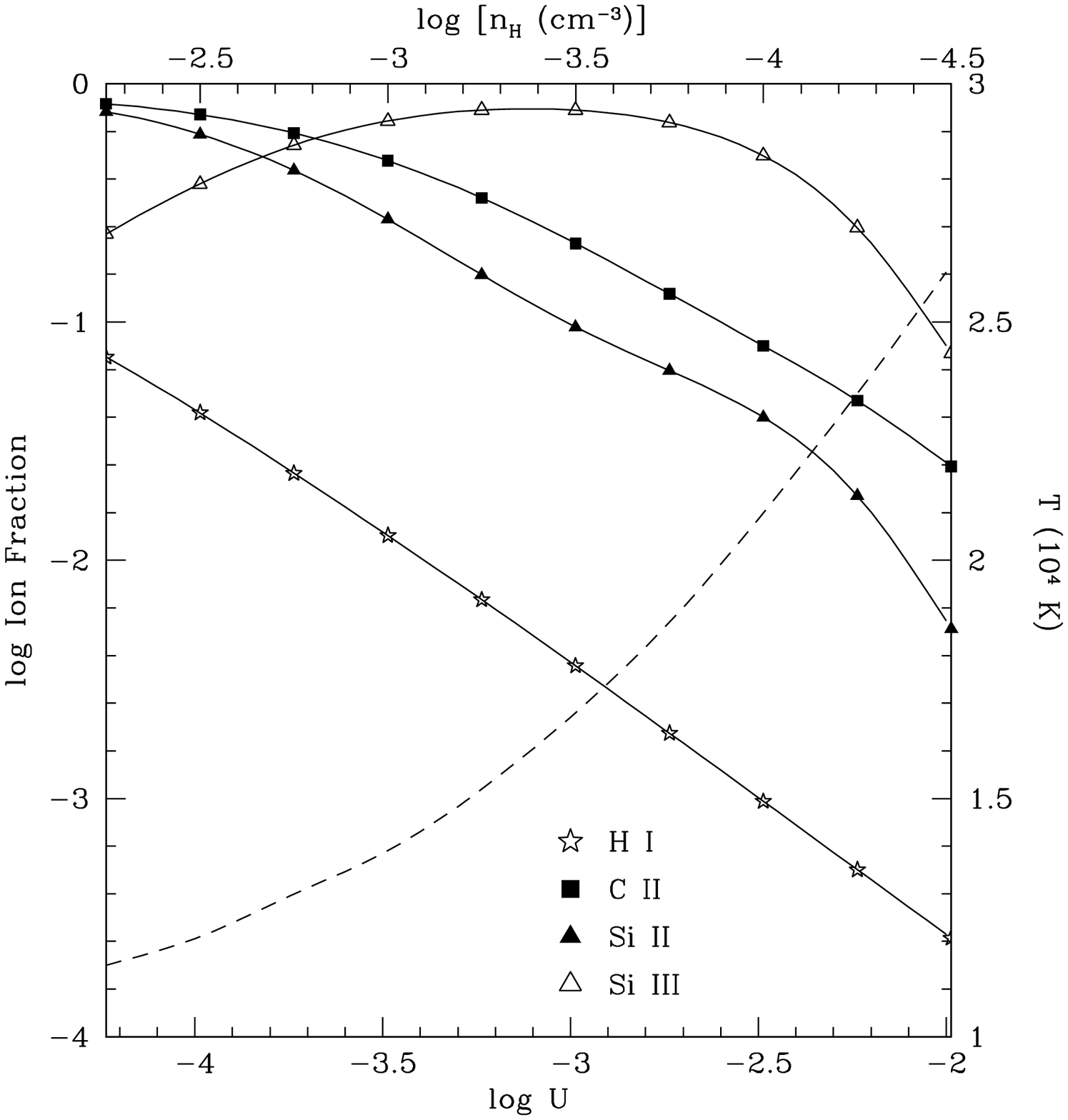}
\caption[]{Logarithmic ion fraction (e.g. 
\ion{Si}{2}/Si$_{\rm total}$) from the model shown in 
Figure~\ref{model} vs. log $U$ (bottom axis) and log 
$n_{\rm H}$ (top axis) for \ion{H}{1} (stars), \ion{C}{2} 
(filled squares), \ion{Si}{2} (filled triangles), and 
\ion{Si}{3} (open triangles). The temperature of the gas is 
also shown with a dashed line using the linear scale on the 
right-hand axis.\label{ionfracs}}
\end{figure}

Even if the absorber is a multiphase medium, we can still 
derive useful estimates of the heavy element abundances 
because the \ion{H}{1}, \ion{C}{2}, and \ion{Si}{2} 
absorption should predominantly originate in the same, 
lower-ionization, phase. To show this, we plot in 
Figure~\ref{ionfracs} the ion fractions of \ion{H}{1}, 
\ion{C}{2}, \ion{Si}{2}, and \ion{Si}{3} vs. log $U$ from 
the same model shown in Figure~\ref{model}. We also show 
the gas temperature predicted by the model over this range 
of $U$. From this figure one can see that the ion fractions 
of \ion{H}{1}, \ion{C}{2}, and \ion{Si}{2} show very 
similar trends as the gas becomes more ionized, and it is 
reasonable to assume that the lines of these three species 
arise in the same gas. Furthermore, \ion{Si}{3} does not 
follow the lower ionization species, and it would not be 
surprising to find that there is somewhat more ionized gas 
that produces a \ion{Si}{3} line that is substantially 
stronger than the \ion{Si}{2} line. The bulk of the 
\ion{H}{1} detected in higher Lyman series lines should be 
affiliated with the \ion{Si}{2}/\ion{C}{2} gas. However, 
the higher-ionization \ion{Si}{3}-bearing gas may produce 
weaker \ion{H}{1} absorption that may be detectable in the 
\lya transition (unless it is confined to the strongly 
saturated portion of the \lya profile). More importantly, 
it is evident from Figure~\ref{ionfracs} that some 
\ion{Si}{3} arising in the \ion{Si}{2}/\ion{C}{2} phase is 
likely to be detectable, and this may explain the 
indications of a weaker component in the \ion{Si}{3} 
profile at the velocity of the \ion{Si}{2} and \ion{C}{2} 
lines. The column density of this weaker component is 
poorly constrained due to blending of the two components 
and the S/N of the data, but fitting two components to the 
\ion{Si}{3} profile with the Fitzpatrick \& Spitzer (1997) 
profile-fitting code and the line spread functions from the 
STIS Handbook (Leitherer et al. 2001) yields log 
$N$(\ion{Si}{3}) $\approx 11.9\pm 0.2$ for the weaker 
component.

%\clearpage

Assuming the \ion{H}{1}, \ion{C}{2}, and \ion{Si}{2} lines 
go together and adopting log $N$(\ion{Si}{3}) $\approx 
11.9\pm 0.2$ for the \ion{Si}{2}/\ion{C}{2}-bearing gas, we 
find from the photoionization models that the following 
abundances are consistent with the observed column 
densities within the $1\sigma$ uncertainties:

\begin{equation}
[{\rm C/H}] = -1.2^{+0.3}_{-0.2},
\end{equation}
and
\begin{equation}
[{\rm Si/C}] = +0.2\pm 0.1,
\end{equation}
with
\begin{equation}
{\rm log} \ n_{\rm H} = -2.8\pm 0.3.
\end{equation}
At the current S/N level, the presence of a more highly 
ionized phase that produces an additional \ion{Si}{3} 
component (discussed in the previous paragraph) does not 
conflict with the upper limits on \ion{Si}{4} and 
\ion{C}{4}. However, at higher S/N levels, \ion{C}{4} 
and/or \ion{C}{2} absorption associated with this more 
highly ionized component should be detectable. It is 
possible that some of the \ion{C}{2} absorption that we 
have already detected is associated with the more highly 
ionized extra \ion{Si}{3} component. If so, then we have 
overestimated [C/H] and we have underestimated [Si/C]. 
Decreasing [C/H] and increasing [Si/C] to compensate for 
this would only strengthen our interpretation of the nature 
of this absorber (\S 4).

The model has some interesting implications. With these 
constraints on $n_{\rm H}$, the model predicts $11,000 
\lesssim T \lesssim 14,000$ K (see Figure~\ref{ionfracs}), 
and the cloud pressure $P/k \approx 40$ K cm$^{-3}$. 
However, the thickness of the gas slab is only $\sim$70 pc 
for log $n_{\rm H} = -2.8$. This is vastly smaller than 
expected for a gravitationally confined \lya cloud with an 
ordinary ratio of baryons to dark matter; a photoionized, 
self-gravitating cloud with log $N$(\ion{H}{1}) = 15.85, 
the inferred temperature, and a gas fraction $f_{g} \approx 
\Omega _{b}/\Omega _{m}$ would be several orders of 
magnitude larger (see Schaye 2001). The inferred absorber 
thickness is somewhat sensitive to the ionization 
parameter. For example, if we decrease the density to log 
$n_{\rm H}$ = $-3.1$ (i.e., log $U = -3.35$), then the 
thickness increases to $\sim 350$ pc.  This is still much 
smaller than expected for a self-gravitating cloud. The 
cloud mass is uncertain due to the unknown geometry and 
impact parameter. If spherical, the cloud has a rather 
small baryonic mass on the order of $10 - 400$ M$_{\odot}$, 
but the mass could be substantially larger with other 
geometries. Similar properties have recently been derived 
for a subset of weak \ion{Mg}{2} systems at $z \sim$ 1 
(Rigby, Charlton, \& Churchill 2002).

The small line-of-sight size implied by the model indicates 
that either (1) the fraction of the absorber mass in the 
form of baryonic gas is extremely small ($f_{g} \sim 10^{-
6} - 10^{-5}$), (2) the absorber is pressure-confined by an 
external medium, or (3) the gas is out of equilibrium and 
will rapidly expand or evaporate. A small $f_{\rm g}$ could 
arise in a small dark matter halo that expelled most of its 
gas in an early episode of star formation. Mac Low \& 
Ferrara (1999) have shown that SN can expel most of the ISM 
from a very low-mass object. Alternatively, the baryons may 
have been photoevaporated from a small dark matter halo 
during the reionization epoch (e.g., Klypin et al. 1999; 
Barkana \& Loeb 1999). Assuming the current gas mass is 
$\sim 10$ M$_{\odot}$, the required $f_{g}$ would imply 
that the mass of the dark matter halo is $10^{6} - 10^{7}$ 
M$_{\odot}$. A substantial fraction of the gas can be 
removed from a dark matter halo in this mass range (see 
references above). However, the baryon removal process must 
be remarkably efficient, and the gas which remains in the 
dark matter halo must be in a relatively quiescent state 
(there is some evidence of component structure in the 
absorption lines, but the profiles are quite simple 
compared to many QSO absorption systems). Furthermore, the 
fact that the baryonic mass (and therefore the implied dark 
matter halo mass) could be substantially higher with other 
(perhaps more likely) geometries such as a filament or 
sheet makes baryon evacuation more difficult and the 
required $f_{g}$ value more improbable.

\begin{figure}
\plotone{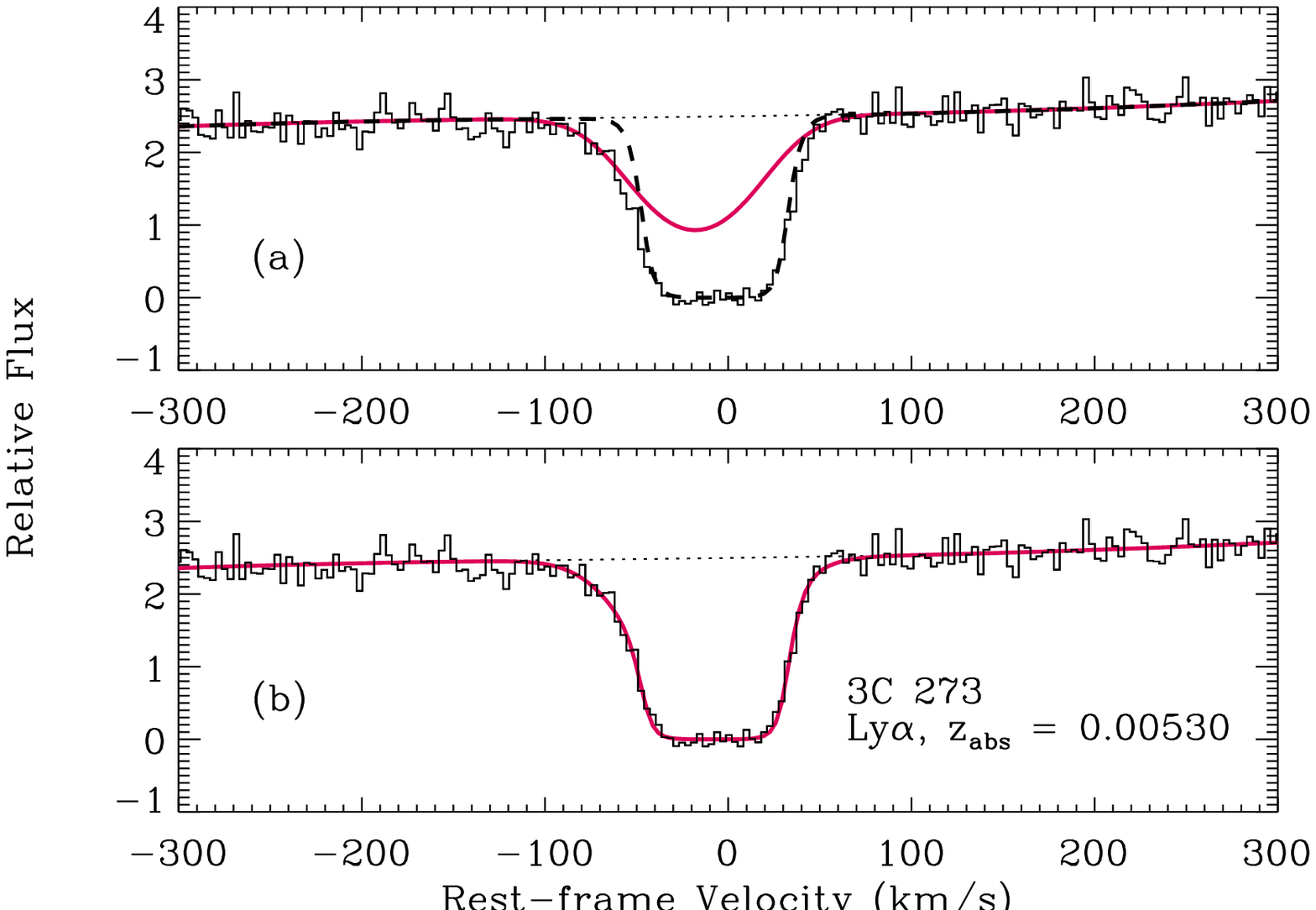}
\caption[]{Broad-component fit to the \lya line at \zabs\ = 
0.00530. In both panels, the observed \lya profile is shown 
with a histogram plotted versus rest-frame velocity ($v = 
0$ \kms at \zabs\ = 0.00530). In panel (a), a broad 
component with $b$ = 41 \kms\ and log $N$(\ion{H}{1}) = 
13.73 is overplotted on the observed profile with a thick 
solid line, and the main component with $b$ = 16.1 \kms\ 
and log $N$(\ion{H}{1}) = 15.85 (as derived by Sembach et 
al. 2001 from a curve-of-growth analysis) is shown with a 
thick dashed line. Panel (b) shows a two-component fit to 
the \lya line including both components shown in panel 
(a).\label{lya005}}
\end{figure}

It seems more likely that the absorber is pressure confined 
by an external medium. For purposes of illustration, 
Figure~\ref{lya005} shows how the \lya profile at \zabs\ = 
0.00530 can easily accomodate a broad component with a $b-
$value as large as 41 \kms . If due to the external 
confining medium, this broad-component Doppler parameter 
implies that $T_{\rm ext} \leq 1 \times 10^{5}$ K, and 
therefore the external medium would need a density $n_{\rm 
ext} \gtrsim 4 \times 10^{-4}$ cm$^{-3}$ to confine the 
metal-bearing gas with the pressure derived above. This is 
a plausible density for a galaxy halo or the intracluster 
medium in a galaxy group. However, the fit shown in 
Figure~\ref{lya005} is not unique, and it is possible that 
a substantially broader \ion{H}{1} component is present. A 
sufficiently broad and weak component could be hidden in 
the noise, particularly since the strength of such a 
feature could be inadvertently reduced by the continuum 
fitting process. Consequently, $T_{\rm ext}$ could be 
substantially higher and $n_{\rm ext}$ accordingly lower.

\subsection{\zabs\ = 0.00337}

We now constrain the metallicity of the 3C 273 \lya cloud 
at \zabs\ = 0.00337, first assuming the gas is 
collisionally ionized (\S 3.2.1), and then considering the 
possibility that it is photoionized (\S 3.2.2). The only 
detected species are \ion{H}{1} and \ion{O}{6}. From a 
curve of growth analysis, Sembach et al. (2001) estimate 
that this system has log $N$(\ion{H}{1}) = 14.41$\pm$0.10 
at this redshift. However, the absence of \ion{C}{4} 
provides a useful constraint for ionization models, and 
even with only two detected species, we can still place 
limits on the metallicity of the gas as follows. The 
logarithmic oxygen abundance is given by
\begin{equation}
\left[ \frac{\rm O}{\rm H}\right] = {\rm log}\left( 
\frac{N({\rm O \ VI})}{N({\rm H \ I)}}\right) + {\rm 
log}\left( \frac{f({\rm H \ I})}{f({\rm O \ VI})}\right) - 
{\rm log}\left( \frac{\rm O}{\rm H}\right) _{\odot} 
\label{metlim}
\end{equation}
where $f$ is the ion fraction and (O/H)$_{\odot}$ is the 
solar oxygen abundance (we take log (O/H)$_{\odot} = -3.26$ 
from Holweger 2001). In general, it is usually possible to 
minimize $f$(\ion{H}{1})/$f$(\ion{O}{6}), and this sets a 
lower limit on [O/H] since the other quantities in 
eqn.~\ref{metlim} are measured. In practice, it can be 
difficult to ascertain the $N$(\ion{H}{1}) that should be 
assigned to the \ion{O}{6}-bearing gas; there is evidence 
that some \ion{O}{6} absorbers are multiphase entities 
(e.g., Tripp, Savage, \& Jenkins 2000), and some of the 
\ion{H}{1} absorption may arise in a 
lower-ionization phase which does not contain \ion{O}{6}.

\subsubsection{Collisional Ionization}

\begin{figure}
\plotone{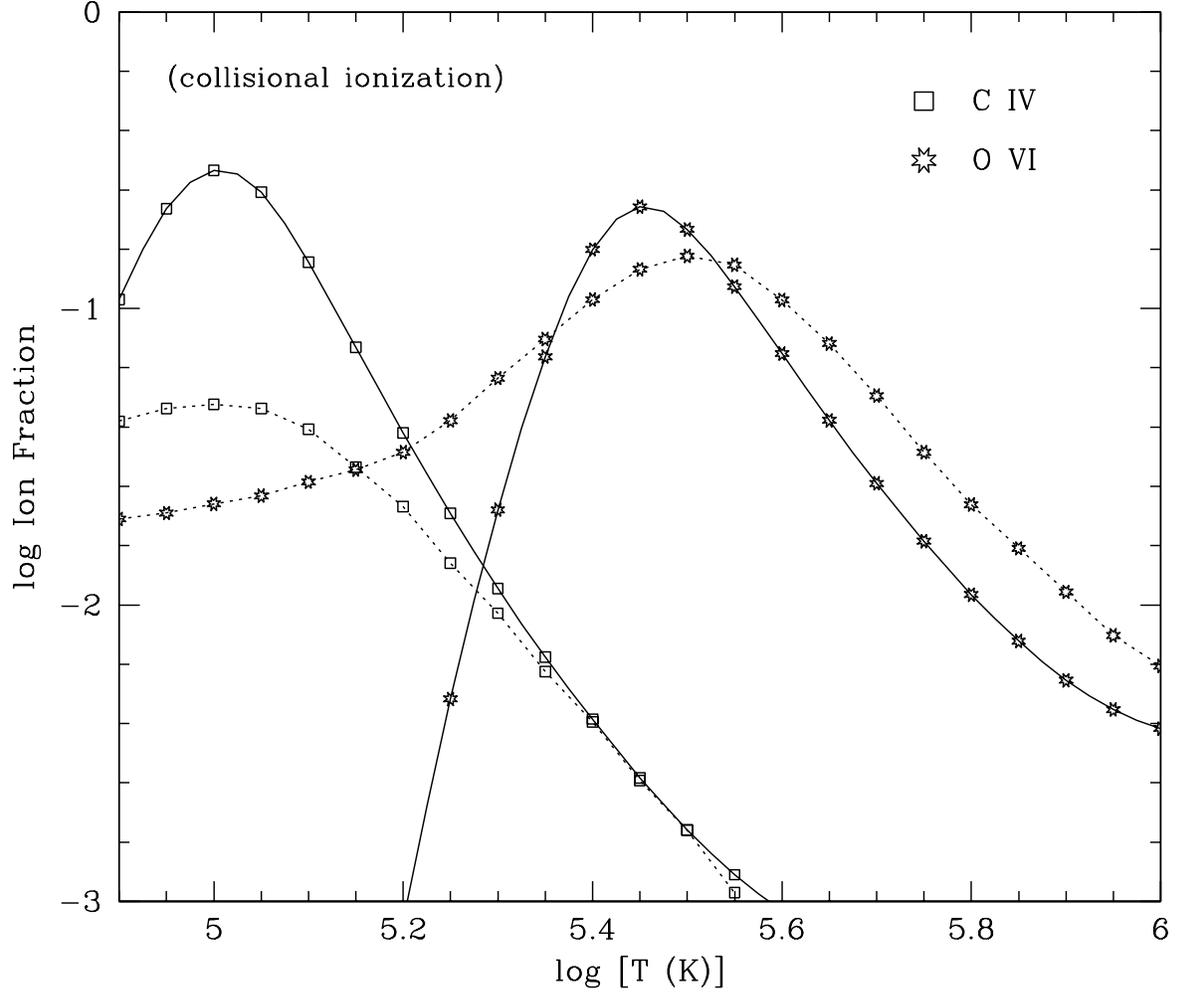}
\caption[]{Ion fractions of \ion{C}{4} (squares) and 
\ion{O}{6} (stars) as a function of temperature for gas in 
collisional ionization equilibrium (solid lines, from 
Sutherland \& Dopita 1993) and nonequilibrium raditiatively 
cooling gas (dashed lines, from Shapiro \& Moore 1976). In 
the nonequilibrium calculation, the gas is initially in 
equilibrium at $10^{6}$ K and then cools 
isochorically.\label{c4o6}}
\end{figure}

We first consider collisional ionization. If the 
\ion{O}{6}-bearing gas is collisionally ionized, then the 
lower limit on \ion{O}{6}/\ion{C}{4} at \zabs\ = 0.00337 
provides a useful lower limit on the gas temperature. 
Figure~\ref{c4o6} shows the \ion{C}{4} and \ion{O}{6} ion 
fractions in collisional ionization equilibrium (Sutherland 
\& Dopita 1993) and in nonequilibrium radiatively cooling 
gas (Shapiro \& Moore 1976) for $T \geq 10^{5}$ K. We see 
that at $T \lesssim 10^{5.2}$ K, the ionization of the gas 
strongly favors \ion{C}{4} if the gas is in equilibrium; in 
this case the \ion{C}{4} ion fraction is one to several 
orders of magnitude higher than the \ion{O}{6} ion 
fraction. To set a lower limit on $T$ using the lower limit 
on \ion{O}{6}/\ion{C}{4} observed at \zabs\ = 0.00337, we 
must make a further assumption about the intrinsic (O/C) 
abundance in this absorber. In general, we use the solar 
relative abundances of Holweger (2001). However, we 
presented evidence in the previous section that 
$\alpha -$elements are overabundant with respect to carbon 
in at least one Virgo \lya cloud. Since O usually follows 
the $\alpha -$element abundance patterns (McWilliam 1997), 
we will assume [O/C] = 0.3; this will set a more 
conservative lower limit on $T$. In this way, we derive $T 
\geq 10^{5.29}$ K from the \ion{O}{6}/\ion{C}{4} limit for 
the equilibrium case. \ion{C}{4} is less strongly favored 
at lower temperatures in the nonequilibrium model, but 
nevertheless the \ion{O}{6}/\ion{C}{4} ratio requires a 
similar lower limit on $T$; from the nonequilibrium model 
we obtain $T \geq 10^{5.18}$.\footnote{We note that in the 
nonequilibrium radiatively cooling gas calculations of 
Schmutzler \& Tscharnuter (1993), the C~IV ion 
fractions are substantially lower, and the O~VI/C~IV 
constraint can be satisfied in much 
cooler gas. The difference is evidently due to atomic data 
since a comparison of the {\it equilibrium} calculations of 
Shapiro \& Moore (1976) to Schmutzler \& Tscharnuter shows 
the same C~IV discrepancy. We favor Shapiro \& Moore 
since the C~IV ion fractions from the large majority 
of collisional ionization equilibrium calculations (e.g., 
Arnaud \& Rothenflug 1985; Shull \& van Steenberg 1982; 
Sutherland \& Dopita 1993; Mazzotta et al. 1998) agree with 
the Shapiro \& Moore steady state results.}

To constrain the \ion{H}{1} column associated with the 
\ion{O}{6} gas, we turn to the \lya profile. If the 
\ion{O}{6} is due to hot, collisionally ionized gas, then 
the corresponding \ion{H}{1} absorption line should be 
broad. We note that since the observed 
\ion{O}{6}/\ion{C}{4} limit requires a similar temperature 
in the equilibrium and non-equilibrium radiatively cooling 
gas models, in this absorber a broad component is expected 
even if the gas was suddenly shock-heated and subsequently 
cooled faster than it could recombine (as modeled by 
Shapiro \& Moore 1976). The \ion{H}{1} $b-$value derived by 
Sembach et al. (2001) from the COG analysis, $b = 
30^{+4.3}_{-3.8}$ \kms , is not compatible with the limits 
on $T$ from \ion{O}{6}/\ion{C}{4}; this $b$ implies a 
temperature upper limit ($T \lesssim 55,000$ K) at which 
\ion{C}{4} should be substantially stronger than \ion{O}{6} 
if collisionally ionized. Even at the 3$\sigma$ upper 
limit, this $b-$value implies that \ion{C}{4} should be 
stronger than or at least comparable to the \ion{O}{6} (see 
Figure~\ref{c4o6}). 

\begin{figure}
\plotone{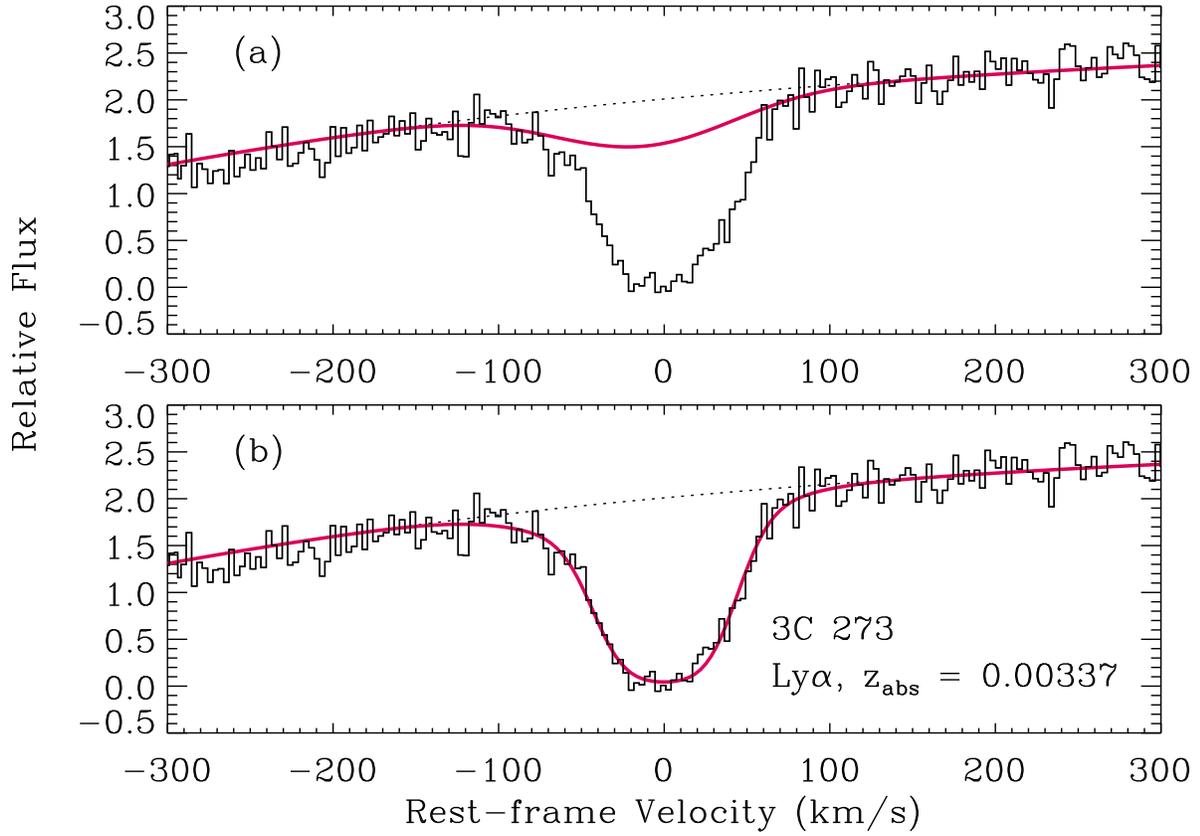}
\caption[]{Broad-component fit to the \lya line at \zabs\ = 
0.00337. As in Figure~\ref{lya005}, the observed \lya 
profile is shown with a histogram plotted versus rest-frame 
velocity ($v = 0$ \kms at \zabs\ = 0.00337). In panel (a), 
a broad component with $b$ = 72 \kms\ and log 
$N$(\ion{H}{1}) = 13.42 is overplotted on the observed 
profile. Panel (b) shows a two-component fit to the \lya 
line including the broad component shown in panel 
(a).\label{lya003}}
\end{figure}

However, inspection of the top panel of 
Figure~\ref{stack003} reveals evidence of structure in the 
\lya profile. It is apparent that a sufficiently broad 
\ion{H}{1} component associated with the \ion{O}{6} is 
allowed by the data as long as the \lya profile probes 
multiple phases of the gas. To set a lower limit on [O/H] 
at \zabs\ = 0.00337, we must minimize 
$f$(\ion{H}{1})/$f$(\ion{O}{6}). In collisional ionization 
equilibrium, this occurs at log $T \approx$ 5.5, where log 
$f$(\ion{H}{1})/$f$(\ion{O}{6}) = $-5.14$ (Sutherland \& 
Dopita 1993). Figure~\ref{lya003} shows a fit to the \lya 
profile including a broad component consistent with log $T$ 
= 5.5 (which requires $b \approx$ 72 \kms ). The upper 
panel in Figure~\ref{lya003} shows the broad component 
only, with $b$ = 72 \kms\ and log $N$(\ion{H}{1}) = 13.42. 
To obtain a good fit to the overall profile, an additional, 
narrower \ion{H}{1} component due to a separate, cooler 
phase must be added, and the lower panel in 
Figure~\ref{lya003} shows a fit including the broad 
component in Figure~\ref{lya003}a as well as a cooler 
component with $b_{\rm cool}$ = 33 \kms\ and log $N_{\rm 
cool}$(\ion{H}{1}) = 14.20. The two-component model shown 
provides an acceptable fit to the \lya profile and is 
compatible with the Ly$\beta$ profile recorded with {\it 
FUSE}. So, with log $N$(\ion{H}{1}) = 13.42, log 
$f$(\ion{H}{1})/$f$(\ion{O}{6}) $\geq -5.14$, and log 
$N$(\ion{O}{6}) = 13.32 for the hot, \ion{O}{6} gas, we 
obtain [O/H] $\geq -2.0$.

Figure~\ref{c4o6} compares the equilibrium \ion{O}{6} and 
\ion{C}{4} ion fractions to those from a particular 
non-equilibrium model. Several other 
non-equilibrium collisionally ionized gas models have been 
developed such as conductive interfaces or turbulent mixing 
layers between very hot gas and cool clouds (see \S 8 in 
Sembach et al. 1997 for a brief review). Some of these 
other models, e.g., the conductive interface model of 
Borkowski et al. (1990), can satisfy the 
\ion{O}{6}/\ion{C}{4} limit in the \zabs\ = 0.00337 Virgo 
absorber. However, we note that the turbulent mixing layer 
model of Slavin et al. (1993) does not produce enough 
\ion{O}{6}; all of the parameter combinations considered by 
Slavin et al. lead to \ion{O}{6}/\ion{C}{4} $<$ 1. 
Moreover, the \ion{O}{6} column densities from a single 
turbulent mixing layer are more than an order of magnitude 
smaller than the observed \ion{O}{6} column in this system, 
so a large number of layers would be needed to explain the 
observations.

\subsubsection{Photoionization}

We next derive constraints on the metallicity of the \zabs\ 
= 0.00337 system assuming the \ion{O}{6} is photoionized. 
In this case, we assume that all of the neutral hydrogen 
derived from the curve-of-growth analysis (log 
$N$(\ion{H}{1}) = 14.41) is associated with the \ion{O}{6} 
phase since this provides the most conservative lower limit 
on [O/H]. In this circumstance, the 
\ion{O}{6}-bearing gas is consistent with the narrow core 
of the \lya line because it can be much cooler than in the 
collisionally ionized case. We use the photoionization 
models from \S 3.1 to determine the minimum value of 
$f$(\ion{H}{1})/$f$(\ion{O}{6}), but we impose the 
additional stability requirement that the size of the 
absorbing cloud not exceed the Jeans length.\footnote{This 
stability requirement is easily satisfied in the 
best-fitting models in \S 3.1.} With these assumptions, we 
find log $f$(\ion{H}{1})/$f$(\ion{O}{6}) $\gtrsim -4.12$ 
with log $n_{\rm H} \gtrsim -5.5$ and absorber thickness $L 
\lesssim 1.5$ Mpc. Therefore we obtain a similar lower 
limit to the oxygen abundance, [O/H] $\gtrsim -2.0$. A 
smaller $L$ requires a higher metallicity; $L \leq 450$ kpc 
requires [O/H] $\geq -1.5$, for example. In principle, the 
absence of \ion{C}{4} provides an upper bound on the 
metallicity, but this is compromised by the uncertain 
$N$(\ion{H}{1}) associated with the \ion{O}{6}: if the 
absorber is multiphase and log $N$(\ion{H}{1})$_{\rm O VI} 
<$ 14.41, then this increases the lower limit on [O/H] (see 
eqn.~\ref{metlim}). Consequently, we can place a firm lower 
limit on the oxygen abundance but we cannot place a strong 
upper limit on [O/H].

\section{Discussion}

We now offer some comments on the implications of these 
abundance measurements. We first discuss the \zabs\ = 
0.00530 absorber. Using the abundance patterns observed in 
the Milky Way ISM as a reference, we consider whether the 
Si overabundance in this system could be due to depletion 
by dust (\S 4.1) or an intrinsic $\alpha -$element 
overabundance (\S 4.2); we favor the latter. We then 
discuss the environment in which the absorber is found and 
the nature of the absorption system (\S 4.3). Finally, we 
make some brief comments on the detection of \ion{O}{6} at 
the redshift of the Virgo supercluster (\S 4.4).

\subsection{Depletion of C and Si by Dust}

We have obtained {\it gas-phase} abundance measurements for 
carbon and silicon at \zabs\ = 0.00530. In the ISM of the 
Milky Way, both of these elements can be depleted by 
incorporation into dust grains. Therefore the total 
abundances, i.e., gas-phase + 
dust-phase, could be higher than the numbers reported in \S 
3. In the Galactic ISM, Si is usually more substantially 
depleted than C, which would leave little room for dust 
depletion in the Virgo absorber studied here. For example, 
toward $\zeta$ Oph, [C/H] = $-0.4$ while [Si/H] = $-1.3$ 
(e.g., Sofia, Cardelli, \& Savage 1994; Snow \& Witt 1996). 
Before proceeding with this comparison, we note an 
important caveat. In \S 3, we modeled the ionization of the 
Virgo \lya absorber in order to derive the Si and C 
abundances in the gas. In the Milky Way ISM, such modeling 
is usually not applied. Instead, measurements of the column 
densities of ionization stages that are dominant in 
\ion{H}{1} regions are taken to provide a good estimate of 
the elemental abundance, e.g., [C/H] = log 
$N$(\ion{C}{2})/$N$(\ion{H}{1}) $-$ log (C/H)$_{\odot}$. In 
this way, gas-phase abundances of species such as Si, Cr, 
and Fe are found to be substantially reduced compared to 
abundances of elements such as O, S, and Zn, species that 
are depleted lightly or not at all. This is usually assumed 
to be primarily due to depletion by dust. However, in some 
cases ionization effects may also play a role (e.g., 
Jenkins et al. 1998; Sofia \& Jenkins 1998; Jenkins et al. 
2000). In some cases, neutrals are included in the 
abundance estimate (e.g., both \ion{C}{1} and \ion{C}{2} 
are measured for the purpose of estimating [C/H]), but a 
larger error can result from the fact that $N$(\ion{H}{1}) 
can underestimate the total hydrogen density associated 
with a particular heavy element because the dominant 
ionization stage can exist in gas where the H is largely 
ionized. Sembach et al. (2000) have modeled the ionization 
of the ``warm ionized medium'', but ionization corrections 
are often neglected in the ISM 
literature.\footnote{Ionization corrections are most 
substantial in low density clouds; in many Milky Way clouds 
the densities are high enough so that the corrections are 
likely to be small.} With this in mind, we note that in 
interstellar gas in the disk, [Si/H] ranges from $-0.35$ to 
$-1.31$ (Sembach \& Savage 1996) while Sofia et al. (1998) 
find [C/H] $\approx -0.45$ over a wide range of physical 
conditions. In the Galactic halo, Si depletion is reduced, 
e.g., Sembach \& Savage (1996) find that [Si/H] ranges from 
$-0.09$ to $-0.47$ on halo sight lines. Depletion of carbon 
in halo gas has not been extensively studied.

The Galactic carbon and silicon abundances reported by 
Sofia et al. (1998) and Sembach \& Savage (1996), 
respectively, are not derived from the same interstellar 
clouds, and it is not entirely clear how to combine the two 
samples to examine the [Si/C] abundance patterns that might 
result from depletion by dust. There are concerns about 
estimating [Si/C] trends from Si and C measurements in 
different clouds. For example, if dust destruction 
liberates silicon leading to the generally larger Si 
abundances in halo gas (see above), the same processes 
could increase the C abundances in the halo as well. Also, 
the overall metallicity of the ISM might not be constant, 
in which case the [Si/C] ranges derived by combining the 
results of Sofia et al. (1998) and Sembach \& Savage (1996) 
could be misleading. To circumvent these issues, we have 
assembled in Table~\ref{mwism} a sample of interstellar 
clouds in which both \ion{Si}{2} and \ion{C}{2} have been 
reliably measured {\it in the same cloud}, including new 
measurements for two sight lines (see Appendix). We have 
limited this sample to measurements made with the highest 
resolution modes of GHRS and STIS (i.e., FWHM $\leq$ 3.5 
\kms ) in order to minimize potential confusion due to 
unresolved saturation. From this table, we see that the 
majority of the clouds in this sample have [Si/C] $\ll$ 
0.0. In four cases in Table~\ref{mwism}, [Si/C] is 
consistent with the solar ratio or somewhat larger. 
However, in two of these four cases, the \ion{C}{2} line 
approaches zero intensity in the line core, and 
$N$(\ion{C}{2}) could be underestimated due to unresolved 
saturation, as noted by Lehner et al. (2001) [all of the 
\ion{Si}{2} measurements are unlikely to be underestimated 
due to saturation]. Increasing $N$(\ion{C}{2}) would 
decrease [Si/C]. The remaining two cases with high [Si/C] 
values (both from the DI 1388 sight line) are evidently 
uncertain; for these clouds Lehner et al. report rough 
measurements without error bars. It is also worth noting 
that there are tentative indications of $\alpha -$element 
overabundances in some high-velocity clouds (e.g., Richter 
et al. 2001), and the high values of [Si/C] in the HVCs 
toward DI 1388 may be due to intrinsic Si overabundances. 
We conclude that the sample in Table~\ref{mwism} presents 
no compelling evidence that dust depletion could lead to a 
gas-phase silicon overabundance with respect to carbon. On 
the contrary, it seems likely that dust would produce an 
underabundance of Si in the gas phase.

\begin{deluxetable}{ccccclcc}
\footnotesize
\tablewidth{0pc}
\tablecaption{High-Resolution [Si/C] measurements in the 
Milky Way ISM\label{mwism}}
\tablehead{Sight Line & $l$ & $b$ & $d$\tablenotemark{a} & 
Cloud $v_{\odot}$ & \ \ [Si/C]\tablenotemark{b} & 
Comment\tablenotemark{c} & Reference\tablenotemark{d} \\
 \ & (deg.) & (deg.) & (pc) & (km s$^{-1}$) & \ & \ & \ }
\startdata
$\epsilon$ CMa........... & 240 & $-11$ & 130 & $-10$ & $-
0.5\pm 0.1$ & L & 1 \\
$\zeta$ Oph........... & 6 & 24 & 138 & $-15$ & $-
0.8^{+0.2}_{-0.3}$ & D & 2,3 \\
23 Ori........... & 199 & $-18$ & 295 & $-108$ & $-
0.8^{+0.1}_{-0.2}$ & F,O & 4 \\
 \     & \   & \     & \   & $-101$ & $-0.6^{+0.1}_{-0.2}$ 
& F,O & 4 \\
\     & \   & \     & \   & $-93$ & $-0.4\pm 0.1$ & F,O & 4 
\\
\     & \   & \     & \   & $-83$ & $-0.5\pm 0.1$ & F,O & 4 
\\
\     & \   & \     & \   & $-43$ & $-0.5^{+0.1}_{-0.2}$ & 
F,O & 4 \\
$\mu$ Col............ & 237 & $-27$ & 400 & $-9$ & $-
0.7^{+0.2}_{-0.3}$ & F & 5 \\
  \       & \   & \   & \   & 61 & $\leq +0.2^{+0.1}_{-
0.2}$\tablenotemark{e} & F & 5 \\
HD116781A.. & 307 & 0 & 1900 & $-107$ & $-0.5\pm 0.1$ & F & 
6 \\
   \      & \   & \ & \    & $-92$  & $-0.6\pm 0.1$ & F & 6 
\\
HD122879....  & 312 & 2 & 2400 & $-109$ & $-1.2^{+0.3}_{-
0.2}$ & F & 6 \\
DI 1388....... & 291 & $-41$ & MB\tablenotemark{f} & 69 & 
$\leq 0.0^{+0.2}_{-0.3}$\tablenotemark{e} & U & 7 \\
  \    & \   & \   & \                   & 102 & 
+0.2\tablenotemark{g} & U & 7 \\
\    & \   & \   & \                   & 119 & 
0.0\tablenotemark{g} & U & 7
\enddata
\footnotesize
\tablenotetext{a}{Distance to background star (see 
reference in column 8 for source).}
\tablenotetext{b}{All [Si/C] values from the literature 
have been recalculated assuming the Holweger (2001) solar 
abundances. Error bars include uncertainties in the solar 
reference abundances reported by Holweger (2001) as well as 
the uncertainties in the column density measurements.}
\tablenotetext{c}{Comment: L = low density sight line 
within the ``Local Bubble'', D = high density sight line, F 
= forbidden velocity cloud (i.e., velocity is outside of 
the expected range for normal Galactic rotation for this 
direction and background star distance), O = star is behind 
``Orion's Cloak'' (Cowie, Songaila, \& York 1979), U = due 
to the substantial distance of the background star, the 
location of the absorption system is uncertain. Based on 
the observed velocities of Magellanic Bridge gas, these 
absorbers are more likely to be associated with the Milky 
Way than the Magellanic Bridge (Lehner et al. 2001).}
\tablenotetext{d}{Column density measurements from: (1) Gry 
\& Jenkins (2001), (2) Cardelli et al. (1993), (3) Cardelli 
et al. (1994), (4) Welty et al. (1999), (5) Howk, Savage, 
\& Fabian (1999), (6) this paper (see Appendix), (7) 
Lehner, Keenan, \& Sembach (2001).}
\tablenotetext{e}{The C~II absorption approaches the 
zero flux level in the core of the line, and 
$N$(C~II) could be underestimated due to saturation. 
Consequently, these measurements are most conservatively 
treated as upper limits. The upper limits are the nominal 
best values (and corresponding 1$\sigma$ uncertainties) 
from the original sources.}
\tablenotetext{f}{This star is located within the gaseous 
``Magellanic Bridge'' between the Large Magellanic Cloud 
and the Small Magellanic Cloud.}
\tablenotetext{g}{Column density uncertainties are not 
reported in the original reference.}
\end{deluxetable}

\subsection{$\alpha-$Element Abundances}

It is quite possible that silicon is intrinsically 
overabundant in the Virgo \lya cloud at \zabs\ = 0.00530, 
and that dust depletion affects the abundances by only a 
small amount if at all. This interpretation has interesting 
implications. It is well known that the $\alpha-$group 
elements, including silicon, are overabundant relative to 
iron by roughly 0.3 dex in low metallicity stars in various 
locations in the Milky Way (Wheeler et al. 1989; McWilliam 
1997, and references therein). Following Tinsley (1979), 
this is generally believed to indicate that the heavy 
element enrichment is dominated by the output from Type II 
supernovae (SN II). The nucleosynthetic origins of carbon 
are somewhat complex, and carbon abundance trends in stars 
are highly uncertain, but McWilliam (1997) notes that there 
are no compelling indications that C is overabundant 
(relative to Fe) in 
low-metallicity Galactic halo stars. Furthermore, some 
galactic chemical evolution models can produce 
overabundances of Si relative to C, with [Si/C] as high as 
$\sim$0.5 (e.g., Timmes et al. 1995). Therefore there is 
reason to expect [Si/C] $\sim$ 0.3 in low-metallicity 
objects such as this Virgo \lya cloud. 

\subsection{Nature of the Absorber at \zabs\ = 0.00530}

There is also long-standing evidence that $\alpha-$group 
elements are overabundant (with respect to iron) in the 
X-ray emitting hot intracluster medium (ICM) in several 
clusters including Virgo (e.g., Canizares et al. 1982; 
Mushotzky et al. 1996; Matsumoto et al. 1996). This has 
usually been ascribed to galactic winds driven 
predominantly by SN II (e.g., Loewenstein \& Mushotzky 
1996, but see the discussion of caveats in Gibson, 
Loewenstein, \& Mushotzky 1997). More recent analyses have 
concluded that the Fe abundances increase in the central 
regions of several clusters while the $\alpha-$group 
abundances remain more-or-less constant (e.g., Finoguenov, 
David, \& Ponman 2000; Kaastra et al. 2001). This likely 
indicates that ejecta from Type Ia SN contribute to the ICM 
enrichment in the central regions where the cD and other 
old elliptical galaxies are located, but SN II ejecta from 
early ``protogalactic'' winds dominate the enrichment in 
the rest of the ICM. The sight line to 3C 273 is 
$>10.3^{\circ}$ away (i.e., projected distance $\gtrsim 3$ 
Mpc) from M87 and the X-ray emitting ICM in the Virgo 
cluster. Consequently, it is interesting that the Virgo 
\lya absorber at \zabs\ = 0.00530 also shows evidence of an 
overabundance of one of the $\alpha$ elements. This may 
indicate that the early SN II winds affect a very broad 
region of the cluster, a region much more extended than the 
easily observed hot gas. This also has implications for the 
issues briefly noted in \S 1. For example, if the 
intergalactic gas within Virgo is enriched by dynamical 
processes such as ram-pressure stripping, the abundance 
patterns should reflect a significant contribution from SN 
Ia (Gunn \& Gott 1972). The observation of SN II abundance 
patterns favors supernova driven winds as the source of 
enrichment. Furthermore, if the \zabs\ = 0.00530 absorber 
is a small blob entrained in the hot gas of a galactic 
wind, then the hotter wind gas could provide the pressure 
required to confine the absorber, as discussed in \S 3.1.

We note that analogous abundance trends have been observed 
in other contexts in low- and high-redshift absorption 
systems. For example, Richter et al. (2001) have measured 
abundances in the high-velocity cloud (HVC) Complex C near 
the Milky Way using absorption lines detected in the 
spectrum of PG1259+593 and a well-constrained COG. They 
report [O/H] = $-1.0^{+0.4}_{-0.3}$, [Si/H] = $-0.9\pm 
0.3$, and [Fe/H] = $-1.3^{+0.2}_{-0.1}$. It is interesting 
to note the overabundance (albeit marginal) of the 
$\alpha-$group elements O and Si with respect to Fe in this 
object which, based on its large angular extent, is 
probably relatively close to the Galaxy ($d \sim$ 10 kpc; 
Wakker et al. 1999). In this case the abundance trends 
could be entirely due to dust, which depletes iron more 
strongly than O and Si and can thereby mimic $\alpha$ 
overabundances. The Virgo absorber at \zabs\ = 0.00530 
could be analogous to this HVC, although its \ion{H}{1} 
column density is substantially lower.

There is also evidence of $\alpha-$group overabundances in 
high-redshift damped \lya absorption systems (e.g., Lu et 
al. 1995, 1996; Prochaska \& Wolfe 1999). In comparisons of 
this Virgo \lya absorber to higher redshift systems, it is 
important to note that low$-z$ and high$-z$ absorbers with 
the same $N$(\ion{H}{1}) are probably not dynamically 
equivalent. In cosmological simulations, a \lya system with 
a given column density at $z \sim$ 0 has the same 
overdensity as a system at $z \sim$ 3 with a larger 
$N$(\ion{H}{1}) by a factor of $\sim$ 50 (Dav\'{e} et al. 
1999). Therefore this Virgo \lya absorber may be more 
analogous to a Lyman limit system at high redshift.

However, damped \lya and Lyman limit absorbers are 
generally thought to be associated with individual 
galaxies. Despite its location within the Virgo 
supercluster (or perhaps because of this), it is difficult 
to attribute the \zabs\ = 0.00530 system to such a galaxy. 
After the Virgo \lya absorbers in the 3C 273 spectrum were 
discovered (Morris et al. 1991; Bahcall et al. 1991), 
several studies of the relationships between these systems 
and galaxies were carried out making use of galaxy redshift 
measurements (Salzer 1992; Morris et al. 1993; Hoffman, 
Lewis, \& Salpeter 1995; Salpeter \& Hoffman 1995; Bowen, 
Blades, \& Pettini 1996; Grogin \& Geller 1998; Impey, 
Petry, \& Flint 1999; Penton, Stocke, \& Shull 2000, 2002), 
deep optical imaging (Morris et al. 1993; Rauch, Weymann, 
\& Morris 1996) and 21cm imaging (van Gorkom et al. 1993). 
Despite these intensive searches with good sensitivity to 
faint and low surface-brightness galaxies, no galaxies have 
been found with impact parameter $\rho$ less than 100 kpc 
near the redshift of the \zabs\ = 0.00530 absorber. Damped 
\lya and Lyman limit absorbers are generally believed to 
have cross sections significantly smaller than 100 kpc 
(e.g., Steidel 1993).

There are some known galaxies near $z$ = 0.00530 with $\rho 
\gg$ 100 kpc (see Table~1 of Salpeter \& Hoffman 1995 as 
well as Grogin, Geller, \& Huchra 1998). For example, as 
noted by Salzer (1992), UGC 7549 (NGC 4420) is within 100 
\kms\ of this absorption system with $\rho = 270$ kpc. UGC 
7549 is notable only because it is the brightest galaxy in 
the vicinity of 3C 273; in other respects it appears to be 
an ordinary spiral galaxy. However, there are at least two 
other galaxies at similar projected distances and velocity 
separations: UGC 7612 ($\rho = 244$ kpc, $\mid \Delta v\mid 
= 80$ \kms ) and the low surface brighness galaxy UGC 7642 
($\rho = 241$ kpc, $\mid \Delta v\mid$ = 50 \kms ). 
\ion{H}{1} 1225+01 is at a somewhat larger, but still 
plausible, velocity separation ($\rho = 167$ kpc, $\mid 
\Delta v\mid = 300$ \kms ). \ion{H}{1} 1225+01 is a large 
cloud of \ion{H}{1} detected in 21cm emission (Giovanelli 
\& Haynes 1989), part of which envelops a small dwarf 
galaxy (Salzer et al. 1991). This object is intriguing 
because the \ion{H}{1} envelope has a complex, 
two-component morphology, and the dwarf galaxy shows 
regions of active star formation (Giovanelli, Williams, \& 
Haynes 1991). This is the type of galaxy that may be able 
drive a wind with which the \zabs\ = 0.00530 system could 
be associated (see Heckman et al. 2001 for an observational 
example of such a wind).\footnote{However, we note that 
mechanisms have been proposed (e.g., Binney 2001) which 
would drive winds from luminous larger galaxies such as UGC 
7549.} Theuns, Mo, \& Schaye (2001) have suggested that 
dwarf galaxy winds ``may break up into clouds, which coast 
to large distances [$\sim$300 kpc] without sweeping up a 
significant fraction of the IGM.''  The small size and mass 
derived for the \zabs\ = 0.00530 \lya cloud is 
qualitatively consistent with this suggestion, and as noted 
above, a small cloud entrained in a wind could be naturally 
pressure confined. However, while it is tempting to assign 
the gas in this absorber to one of these galaxies, we see 
no clear method to determine which one, if any, is the 
correct assignment. Given the complex velocity field within 
the Virgo cluster, it remains possible that some of these 
galaxies are interlopers, i.e., the 
three-dimensional distances from the \lya cloud to the 
galaxies are larger than implied by their velocity 
differences. Furthermore, there is evidence that the two 3C 
273 absorbers in Virgo and their nearest galaxies are part 
of a large-scale filament (Penton et al. 2001), in which 
case the absorption line system may not be associated with 
any particular galaxy but rather is part of the general 
intragroup gas.

Lacking a specific galaxy to which this absorption system 
can clearly be assigned, it is interesting to consider 
whether this Virgo absorber might be a relatively pristine 
gas cloud that has only been enriched by a putative first 
generation of stars, i.e., the ``Population III'' stars 
(e.g., Qian, Sargent, \& Wasserburg 2002). Using the 
Population III yields calculated by Heger \& Woosley 
(2002), Oh et al. (2001) have argued that explosions of the 
first very massive stars could produce overabundances of 
$\alpha-$group elements with respect to carbon, exactly as 
observed in the Virgo \lya cloud. However, Oh et al. (2001) 
have also shown that while there is evidence of Pop III 
enrichment in low-metallicity stars with $Z \lesssim 10^{-
3}Z_{\odot}$, at higher metallicities enrichment from 
normal stars appears to be dominant. Therefore the 
metallicity derived for the Virgo absorber (see \S 3) 
appears to be too high to be consistent with this 
hypothesis.

\subsection{Warm/Hot Gas in Virgo}

In the process of constraining the oxygen abundance in the 
Virgo \ion{O}{6}/\lya absorber at \zabs\ = 0.00337, we have 
explored the physical conditions of this system (\S 3.2). 
We now briefly comment on these physical condition 
constraints. Based on hydrodynamic simulations of 
cosmological structure growth, it has been suggested that 
at the present epoch, 30-50\% of the baryons are in 
low-density, shocked gas at $10^{5} - 10^{7}$ K, the 
so-called ``warm/hot intergalactic medium'' (Cen \& 
Ostriker 1999a; Dav\'{e} et al. 2001). Lithium-like 
\ion{O}{6} is useful for probing such gas since its ion 
fraction is maximized at $T \approx 10^{5.5}$ K in 
collisionally ionized gas (see Figure~\ref{c4o6}), and 
low$-z$ \ion{O}{6} absorbers do harbor a substantial 
quantity of baryons (Tripp, Savage, \& Jenkins 2000; Savage 
et al. 2002). However, the nature of \ion{O}{6} absorbers 
is not yet clear. Are these systems collisionally ionized 
or photoionized? Do they arise in unvirialized, large-scale 
filaments (as predicted by the cosmological simulations, 
see Cen et al. 2001; Fang \& Bryan 2001), higher density 
galaxy groups (Mulchaey et al. 1996), individual galaxies, 
or galactic winds?

The 3C 273 \ion{O}{6}-\lya absorber at \zabs\ = 0.00337 
provides some interesting clues. The absorption line 
properties are consistent with expectations for 
collisionally ionized, warm-hot intergalactic gas at $T > 
10^{5}$ K. As \lya systems go, this absorber is located in 
a region of relatively high galaxy density (Grogin \& 
Geller 1998). Given the low \ion{O}{6} equivalent width, 
this absorber may not be located in the {\it typical} low-
overdensity environment expected based on the cosmological 
simulations (see, e.g., Figure 4 in Cen et al. 2001), but 
this is only one case, and it is quite likely that 
analogous \ion{O}{6} systems can be found in the 
simulations. Like the \zabs\ = 0.00530 system, there are no 
galaxies very near the sight line at \zabs\ = 0.00337 
(Hoffman et al. 1998, and references therein), and it is 
not clear whether or not this absorber is in a virialized 
region. Additional galaxy redshift measurements currently 
underway (K. McLin et al., in preparation) will help 
illuminate the nature of this absorption system. Other 
\ion{O}{6} absorbers at higher redshifts have line 
properties similar to the \zabs\ = 0.00337 system, e.g., 
similar lower limits on \ion{O}{6}/\ion{C}{4} and evidence 
of broad components in the \lya profiles (e.g., Tripp \& 
Savage 2000; Tripp et al. 2001; Savage et al. 2002). It is 
interesting to note that in all of these absorbers, the 
broad \ion{H}{1} lines are juxtaposed with at least one 
narrow \ion{H}{1} line. This might be expected if the 
broad-component gas was 
shock-heated as high-velocity gas collided with the 
lower-ionization gas that gives rise to the observed narrow 
component. On the other hand, if the broad \ion{H}{1} 
component is due to Hubble broadening over a long 
pathlength (as may be required if the \ion{O}{6} is 
photoionized, see Tripp et al. 2001), then there is no 
particular reason to expect the observed correspondence 
between the broad and narrow \ion{H}{1} lines. The 3C 273 
\ion{O}{6} system at \zabs\ = 0.00337 is particularly 
valuable for understanding these absorbers because the low 
redshift allows an intensive investigation of the absorber 
environment, and because the QSO is the brightest known.

\section{Summary}

We have obtained good constraints on the abundances of 
carbon and silicon in a \lya absorption system at \zabs\ = 
0.00530 in the spectrum of 3C 273; this absorber is located 
within the southern extension of the Virgo cluster. Based 
on ionization modeling, we find that the carbon abundance 
in the gas phase is roughly 1/15 of the solar abundance, 
and Si, an $\alpha-$element, is overabundant relative to C 
by $\sim$0.2 dex. The H density is $\sim 0.0015$ cm$^{-3}$, 
the absorber thickness is $\sim$70 pc, and the gas pressure 
$p/k \approx$ 40 cm$^{-3}$ K. The small line-of-sight 
thickness suggests that the absorber is pressure confined 
by an external medium or is out of equilbrium and will soon 
expand or evaporate (a self-gravitating cloud would require 
a remarkably high dark matter to baryon ratio). The Si 
overabundance is probably not due to dust depletion. 
Instead, this likely reflects an intrinsic abundance 
pattern. Overabundances of 
$\alpha-$group elements have been observed in the hot X-ray 
emitting gas in the Virgo cluster, and these abundance 
patterns are believed to be produced by galactic winds 
containing predominantly SN II ejecta. We favor such an 
explanation for this 3C 273 absorber as well because there 
are no galaxies particularly nearby, as might be expected 
in other scenarios, but there are galaxies at projected 
distances of $\sim$200 kpc which could plausibly drive an 
escaping wind. The projected distance from the 3C 273 sight 
line to the region detected in X-rays is $\sim$3 Mpc, and 
therefore the putative SN II winds may be widespread in the 
Virgo cluster. If this result is found for other sight 
lines, SN II winds, perhaps ``protogalactic'' winds, may 
not be unique to the central regions of clusters and may in 
fact be a general phenomenon in galaxies. Detections of 
other heavy elements in this absorber will require 
substantial improvements in the S/N of the spectrum (at the 
same resolution). Perhaps the most promising transition to 
search for in future observations is the strongest 
\ion{Fe}{2} transition at 2382.8 \AA . 

We have also estimated the metallicity of the 
Ly$\alpha$/\ion{O}{6} absorber at \zabs\ = 0.00337 in the 
3C 273 spectrum. We obtain [O/H] $\gtrsim -2.0$ from 
photoionized as well as collisionally ionized models. If 
the gas is collisionally ionized, then the lower limit on 
the \ion{O}{6}/\ion{C}{4} ratio requires $T \gtrsim 
10^{5.3}$ K in the \ion{O}{6} phase.

Several other sight lines to QSOs behind the Virgo 
supercluster have recently been observed or will be 
observed during the next year with the STIS E140M echelle 
mode. For example, it will be useful to examine the sight 
line to RXJ1230.8+0115 (Read, Miller, \& Hasinger 1998), 
which is only 54' from 3C 273. These nearby sight lines 
will likely provide additional valuable constraints on the 
heavy element enrichment of the IGM.

\acknowledgements

We are indebted to Gary Ferland and collaborators for the 
construction of CLOUDY, and for making the code freely 
available. We also thank Jane Charlton, John Mulchaey, Bill Oegerle, Joop 
Schaye, and especially Ken Sembach for helpful comments. 
This research was supported by NASA Grant NAS5-26555. TMT 
acknowledges additional support from NASA LTSA Grant 
NAG5-11136.

\appendix

\section{Measurements of $N$(\ion{C}{2}) and 
$N$(\ion{Si}{2}) in High-Velocity Milky Way Gas in the 
Directions of HD116781A and HD122879}

As discussed in \S 4, there are relatively few Galactic 
interstellar clouds in which the \ion{Si}{2} {\it and} 
\ion{C}{2} column densities have both been reliably 
measured. This is primarily because $N$(\ion{C}{2}) is 
difficult to measure in the Milky Way ISM; the \ion{C}{2} 
resonance transitions are usually very badly saturated, and 
when the spin-changing \ion{C}{2}] $\lambda$2325 line is 
detectable, the corresponding \ion{Si}{2} resonance 
transitions are strongly saturated. There is a 
spin-changing \ion{Si}{2}] transition at 2335 \AA\ that can 
be observed when the \ion{C}{2}] $\lambda$2325 line is 
detected. However, detection of these semiforbidden 
transitions is challenging, and they have mainly been 
observed in sight lines such as $\zeta$ Oph through 
relatively dense clouds. 

The recent high-resolution survey of interstellar 
\ion{C}{1} absorption lines carried out by Jenkins \& Tripp 
(2001) has fortuitously revealed a few additional 
high-velocity interstellar clouds in the spectra of two 
stars, HD116781A and HD122879, in which $N$(\ion{C}{2}) and 
$N$(\ion{Si}{2}) can both be accurately measured. The 
continuum-normalized \ion{C}{2} and \ion{Si}{2} absorption 
profiles of these high-velocity features are shown in 
Figure~\ref{mw116781}. Due to the very high resolution of 
the spectra ($R \approx 200,000$), it is unlikely that 
these high-velocity lines are affected by unresolved 
saturation. Consequently, we can obtain reliable column 
density measurements. In the direction of HD116781A, at 
least two components are readily apparent in the high 
velocity gas. To deblend these components and obtain their 
individual column densities, we have used the profile-
fitting software of Fitzpatrick \& Spitzer (1997) with the 
line-spread function for this slit from the STIS Handbook. 
The heliocentric velocities, Doppler parameters, and column 
densities obtained in this way are summarized in 
Table~\ref{mwlineprop}. Table~\ref{mwism} reports the 
[Si/C] values derived from these column densities. There 
may be evidence of additional components in these profiles, 
but these additional components are not sufficiently 
well-constrained to justify addition of more lines to the 
fits. We note that due to the weakness of the line, the 
wings of the \ion{Si}{2} profile in the HD122879 spectrum 
might be lost in the noise. This could lead to 
underestimates of the $b-$value and $N$(\ion{Si}{2}) in 
this cloud. Based on apparent column density ratios and 
fits to the \ion{Si}{2} line forced to take on larger $b-
$values, we estimate that log $N$(\ion{Si}{2}) might be 
underestimated by 0.2 dex. The error bars in 
Table~\ref{mwism} reflect this source of uncertainty.

A full discussion of the nature of these high-velocity 
features is beyond the scope of this paper. However, a few 
comments are in order. (1) The absorption lines occur at 
velocities well outside the velocity range expected for 
normal Galactic rotation in these directions (see Table 2 
in Jenkins \& Tripp 2001), i.e., these 
are ``forbidden-velocity'' clouds. (2) The [Si/C] 
abundances derived from the two 
high-velocity components toward HD116781A are quite similar 
(and consistent with the same value within the 1$\sigma$ 
uncertainties; see Table~\ref{mwism}), but there are large 
differences in [Si/C] toward HD116781A and HD122879 at 
$v_{\odot} \approx -109$ \kms\ (see Table~\ref{mwism}). It 
seems unlikely that these differences are due to intrinsic 
changes in the interstellar abundances given the spatial 
separation between the sight lines ($\sim$190 pc at the 
estimated distance of HD116781A). As discussed in \S 4, it 
is possible that some of the differences in [Si/C] are 
driven by ionization. The differences between the HD116781A 
and HD122879 sight lines are dramatic; $N$(\ion{C}{2}) is 
only $\sim$0.3 dex smaller toward HD122879, but the 
\ion{Si}{2} column density is reduced by an order of 
magnitude. Based on the modeling of Sembach et al. (2000), 
such large changes cannot be attributed to ionization 
effects if the gas is predominantly photoionized. However, 
this could occur if the gas is predominantly collisionally 
ionized (see Sutherland \& Dopita 1993) and the gas 
temperature exceeds $\sim$15,000 K. This would not be 
surprising since these clouds may well be shock-heated, 
given their large departures from normal Galactic rotation. 
In addition, some of the differences in [Si/C] between the 
sight lines may be due to more substantial dust grain 
destruction in the direction of HD116781A. This could occur 
if silicon is incorporated into more fragile portions of 
dust grains while C is locked into more resilient dust 
components.

\begin{deluxetable}{ccccc}
\tablewidth{0pc}
\tablecaption{Forbidden-Velocity Absorption Lines in the 
Spectra of HD116781A and HD122879\label{mwlineprop}}
\tablehead{Sight Line & Transition & $v_{\odot}$ (\kms 
)\tablenotemark{a,b} & $b$ (\kms )\tablenotemark{a} & log 
$N$\tablenotemark{a} }
\startdata
HD116781A.. & \ion{C}{2} $\lambda 1334.53$ & $-107$ & 
$8.9\pm 0.5$ & 13.68$\pm$0.01 \\
  \       & \ion{Si}{2} $\lambda 1260.42$ & $-109$ & 
$7.0\pm 0.9$ & 12.10$\pm$0.02 \\
  \       & \ion{C}{2} $\lambda 1334.53$ & $-92$  & $5.9\pm 
0.5$ & 13.42$\pm$0.02 \\
  \       & \ion{Si}{2} $\lambda 1260.42$ & $-91$ & $3.6\pm 
0.9$ & 11.74$\pm$0.03 \\
HD122879.... & \ion{C}{2} $\lambda 1334.53$ & $-109$ & 
$7.2\pm 0.2$ & 13.40$\pm$0.01 \\
  \      & \ion{Si}{2} $\lambda 1260.42$ & $-110$ & 
$3.5^{+2.7}_{-1.5}$ & 11.12$\pm$0.08
\enddata
\tablenotetext{a}{Doppler parameter, component velocity, 
and column density measured using the profile-fitting 
software of Fitzpatrick \& Spitzer (1997). Uncertainties in 
this table only include random errors estimated by the 
profile-fitting code; systematic sources of uncertainty 
such as continuum placement are not included in these error 
bars.}
\tablenotetext{b}{For HD116781A, $v_{\rm LSR} = v_{\odot} - 
5.9$ \kms , assuming the standard definition of the Local 
Standard of Rest (Kerr \& Lynden-Bell 1986). For HD122879, 
$v_{\rm LSR} = v_{\odot} - 4.0$ \kms .}
\end{deluxetable}
%\clearpage

\begin{figure}
\plotone{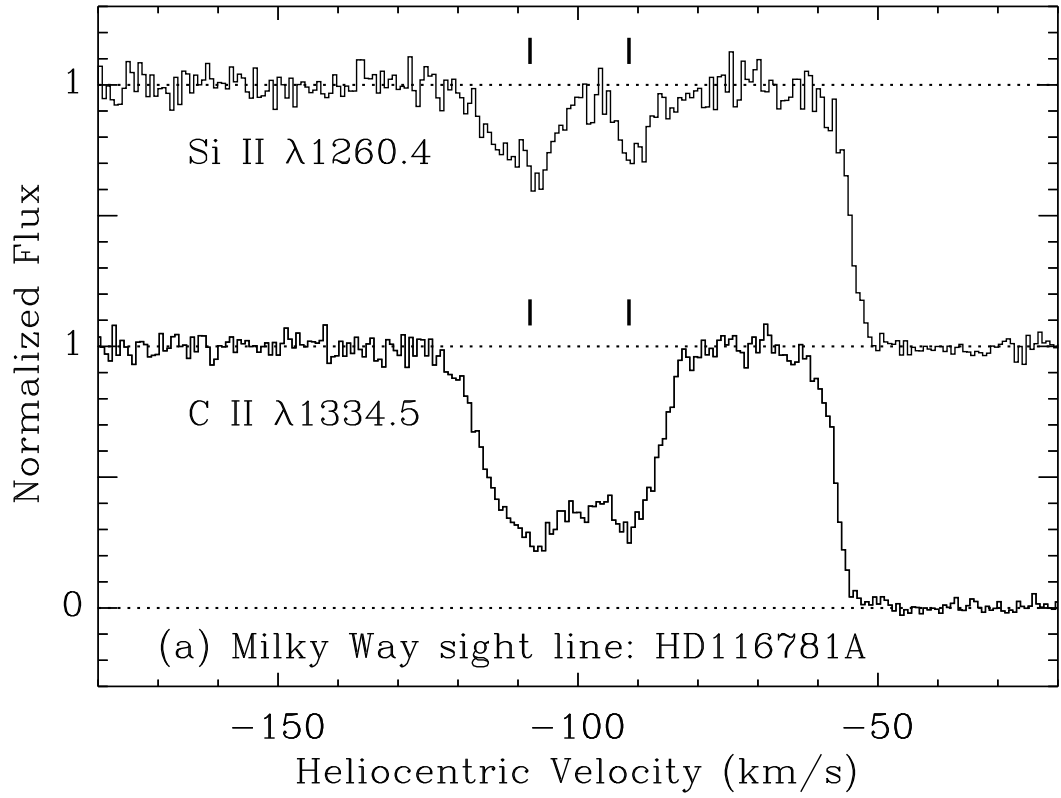}
\caption[]{Continuum-normalized absorption profiles of the 
\ion{C}{2} $\lambda$1334.5 and \ion{Si}{2} $\lambda$1260.4 
lines in the Galactic sight lines towards (a) HD116781A, 
and (b) HD122879, plotted versus heliocentric velocity. The 
components of interest are indicated with tick 
marks.\label{mw116781}}
\end{figure}

\begin{figure}
\figurenum{10}
\plotone{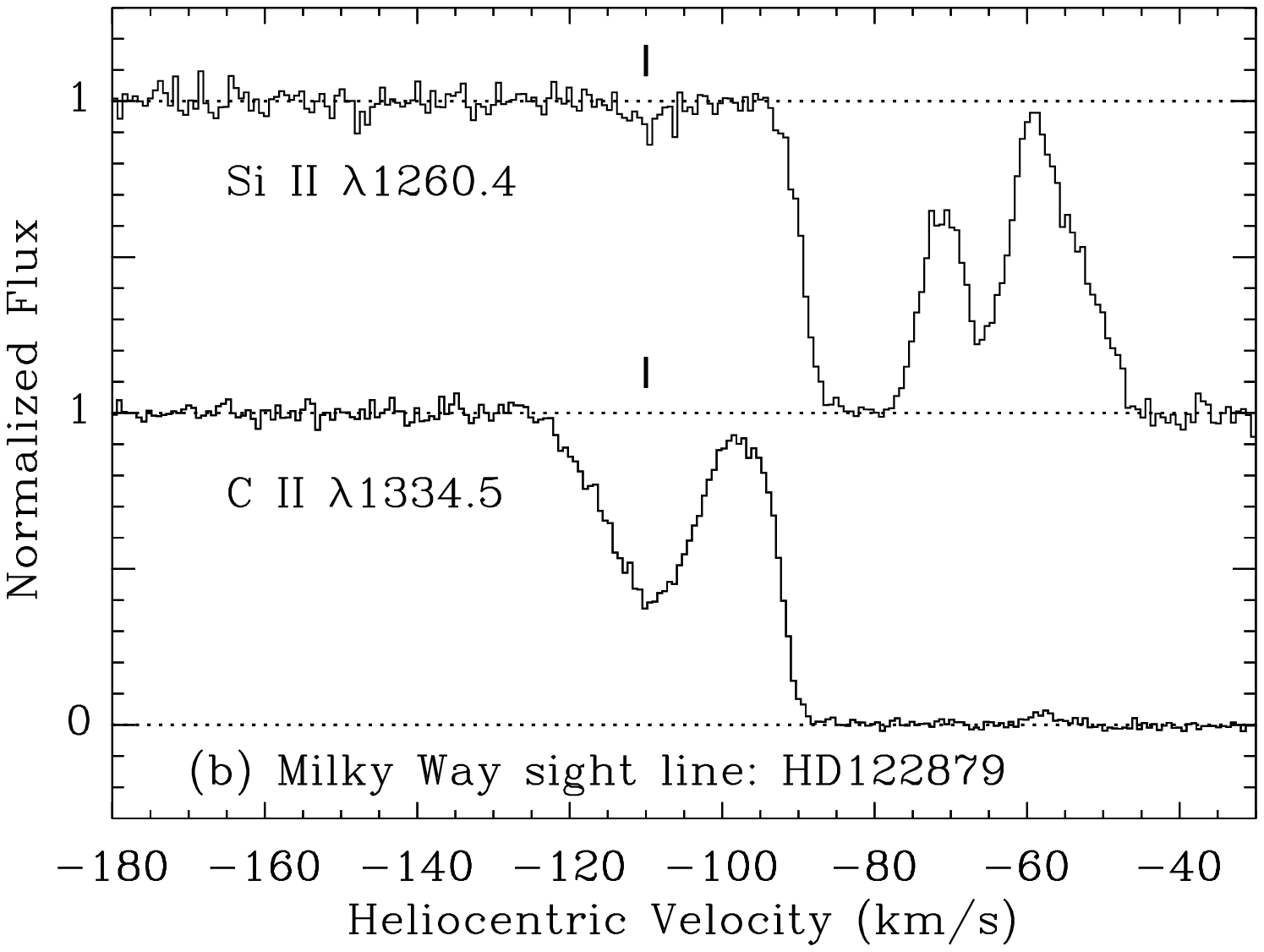}
\caption[]{(continued).}
\end{figure}

\end{document}